\documentclass{article}
\usepackage{ijcai24}

\usepackage{times}
\usepackage{soul}
\usepackage{url}
\usepackage[hidelinks]{hyperref}
\usepackage[utf8]{inputenc}
\usepackage[small]{caption}
\usepackage{graphicx}
\usepackage{amsmath}
\usepackage{amsthm}
\usepackage{booktabs}
\usepackage{algorithm}
\usepackage{algorithmic}
\usepackage[switch]{lineno}

\usepackage{amssymb}
\usepackage{multirow}
\usepackage{makecell} % 表格内换行
\usepackage{pifont}
\usepackage{subcaption}
\title{HyDiscGAN: A Hybrid Distributed cGAN for Audio-Visual Privacy Preservation in Multimodal Sentiment Analysis}

\author{
Zhuojia Wu$^{1}$
\and
Qi Zhang$^{1,4}$\and
Duoqian Miao$^1$\and
Kun Yi$^2$\and
Wei Fan$^3$\And
Liang Hu$^{1,4}$\
\affiliations
$^1$Tongji University\\
$^2$Beijing Institute of Technology\\
$^3$University of Oxford\\
$^4$DeepBlue Academy of Sciences
\emails
\{wuzhuojia, zhangqi\_cs, dqmiao, lianghu\}@tongji.edu.cn,
yikun@bit.edu.cn,
frankfanwei@outlook.com
}
\begin{document}

\maketitle

\begin{abstract}

Multimodal Sentiment Analysis (MSA) aims to identify speakers' sentiment tendencies in multimodal video content, raising serious concerns about privacy risks associated with multimodal data, such as voiceprints and facial images. Recent distributed collaborative learning has been verified as an effective paradigm for privacy preservation in multimodal tasks. However, they often overlook the privacy distinctions among different modalities, struggling to strike a balance between performance and privacy preservation. Consequently, it poses an intriguing question of maximizing multimodal utilization to improve performance while simultaneously protecting necessary modalities. This paper forms the first attempt at modality-specified (i.e., audio and visual) privacy preservation in MSA tasks. We propose a novel \textbf{Hy}brid \textbf{Dis}tributed cross-modality \textbf{cGAN} framework (\textbf{HyDiscGAN}), which learns multimodality alignment to generate fake audio and visual features conditioned on shareable de-identified textual data. The objective is to leverage the fake features to approximate real audio and visual content to guarantee privacy preservation while effectively enhancing performance. Extensive experiments show that compared with the state-of-the-art MSA model, HyDiscGAN can achieve superior or competitive performance while preserving privacy.

\end{abstract}

\section{Introduction}\label{Introduction}
With the growing prevalence of video content on social media, Multimodal Sentiment Analysis (MSA) is poised to provide new opportunities by leveraging multimodal data to enhance and go beyond traditional text-based sentiment analysis~\cite{zhao2023tmmda}. MSA aims to predict the speaker's sentiment by utilizing extra information available in audio and visual content instead of only textual content. The audio/visual modality extracts facial emotions and vocal expressions, enabling a more comprehensive sentiment understanding in wide-ranging applications.%and thereby finding heightened interest and

% By considering multimodal expressions, MSA enables a more comprehensive understanding of sentiment, thereby generating heightened interest and finding wide-ranging applications.

\begin{figure}[t]
\centering
\subfloat[Video data often contains personal information, whereas text content, when properly de-identified, easily maintains privacy.]
% Differences in privacy implications across different modalities, where PII is short for Personally Identifiable Information. When only the de-identified text is available, the speaker's privacy can be effectively preserved.]%Differences in privacy implications across different modalities, where PII is short for Personally Identifiable Information. When only the de-identified text is available, the speaker's privacy can be effectively preserved.
{\includegraphics[width=3.3in]{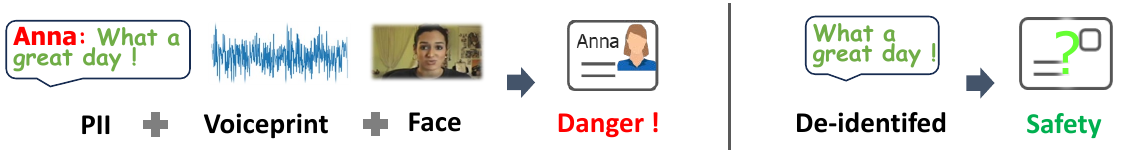}} \\
\subfloat[Differences among our MSA framework, the centralized MSA frameworks, and the DCL-based MSA frameworks.]{\includegraphics[width=3.3in]{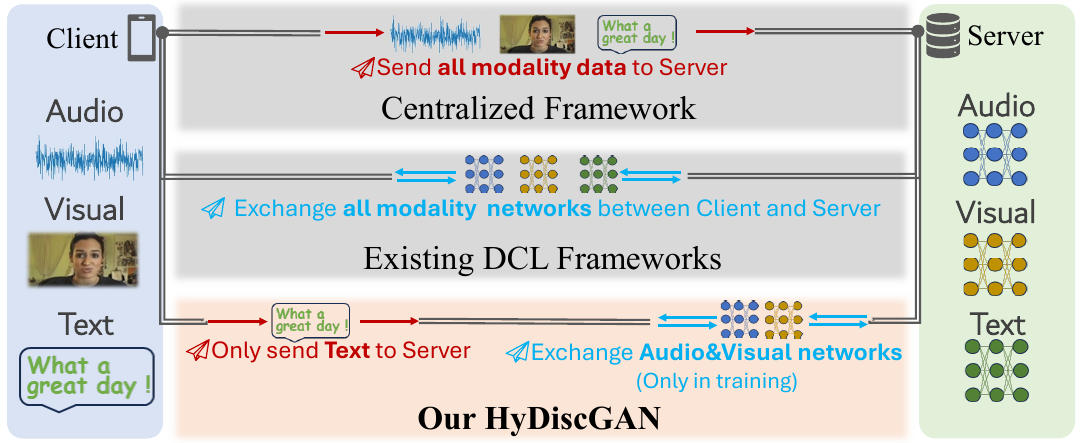}}
\vspace{-2mm}
\caption{Illustration of privacy preservation and MSA frameworks.}
\vspace{-3mm}
\label{figure1}
\end{figure}
Notably, social video data contains massive private information, including Personally Identifiable Information (PII) and biometric data like face images and voiceprints~\cite{regulation2016regulation}. Unfortunately, the misuse of personal information often causes a string of public security incidents, raising widespread concerns regarding personal privacy and security within society~\cite{nguyen2021federated,yang2020federated}. Upon closer examination of video data, we uncover a crucial yet often overlooked fact: \textit{different modalities carry varying requirements for privacy}, as depicted in Figure~\ref{figure1}(a). For instance, legislative efforts aimed at preserving privacy data have emphasized the privacy of personal audio or visual data over textual data~\cite{regulation2016regulation}. In addition, techniques such as introducing noises or blurring faces (e.g., differential privacy~\cite{dwork2006differential}) to de-identify audio and visual data can significantly impede the recognition of sentiment cues. In contrast, de-identifying textual data, such as removing sensitive words, can effectively protect privacy without altering the primary semantics~\cite{WangH023}. The observations encourage us to contemplate an intriguing question: \textit{how to protect the privacy of specific modalities (i.e., audio and visual) when building MSA models?}%regulation2016regulation,de2018guide

Existing MSA approaches typically adopt a centralized paradigm where multimodal data is collected from personal devices and stored centrally for training, achieving excellent performance yet posing considerable challenges and risks in preserving personal privacy, as shown in Figure \ref{figure1}(b). Instead, increasing efforts have been made to apply Distributed Collaborative Learning (DCL) to multimodal tasks~\cite{yu2022multimodal,chen2022fedmsplit}. DCL frameworks~\cite{kairouz2021advances}, such as Federated Learning (FL)~\cite{nguyen2021federated} and Split Learning (SL)~\cite{thapa2022splitfed}, have gained prominence, offering privacy preservation by avoiding centralized data hosting and access. They rely on exchanging multimodal networks between a central server and clients that hold unshareable data for model training and testing, however, they struggle to strive a balance between performance and privacy preservation. Additionally, these endeavors primarily focus on scenarios where all multimodal content is isolated on separate clients, which does not align with the goal of modality-specific privacy preservation in practice. These insights prompt us to explore strategies for \textit{maximizing data utilization to improve performance while simultaneously protecting necessary modalities}. %Consequently, \textit{how to maximize data utilization to enhance performance while protecting necessary modalities} undoubtedly lays a practical breakthrough and focus, yielding significant benefits to DCL-based multimodal tasks.

An intuitive solution to modality-specific privacy preservation is combining centralized and DCL frameworks to create a hybrid distributed learning paradigm, guaranteeing data utilization and protection, respectively. Accordingly, we can initiate a primary idea: train shareable modality data (text) centrally while private modality data (audio and visual) distributively. However, in the setting, where we have separate copies of shareable modality and private modality data on the server and clients respectively, we face a dilemma when it comes to performing model inference. On one hand, performing inference on the server requires access to the data representations of the private modality, which can increase communication costs and pose risks to privacy~\cite{thapa2022splitfed}. On the other hand, performing inference on the client-side necessitates each client to train the entire MSA model to guarantee effective multimodal fusion, needing more client computational resources. Note that it is practically impossible for personal devices (clients) such as smartphones or laptops to have sufficient computing power to accommodate widespread large-scale MSA models. As a result, the hybrid distributed mode inevitably poses two primary challenges stemming from the misaligned treatment between modalities: 1) \textit{achieving effective multimodal alignment}, and 2) \textit{ensuring efficient collaborative communication}.

In light of the above discussion, we propose a novel hybrid distributed collaborative learning framework based on a cross-modality conditional Generative Adversarial Network (cGAN), termed HyDiscGAN. Specifically, we build an audio generator and a visual generator to generate fake features of private audio and visual data respectively in an autoregressive manner. The generators are placed in the server to approximate real features in the clients. On one hand, the generated features are sent to the corresponding audio and visual discriminators in the clients which are regulated by two customized contrastive losses and a cGAN loss. Generators and discriminators are based on Transformers to cater to the sequential audio and visual data. On the other hand, the features are fed into Transformer layers, followed by the gated attention unit to fuse multimodal features of text, visual, and audio, to perform the downstream sentiment analysis. Note that HyDiscGAN is trained in two stages: 1) the cross-modality cGAN is pre-trained to guarantee \textit{effective multimodal alignment}, where global generators and local discriminators are distributively optimized in an alternating manner; 2) the MSA components are trained along with fine-tuning the generators under keeping the discriminators frozen. Its learning process simulates guessing audio and visual (semantic) features conditioned on text inputs, which is inspired by the empirical observation that individuals can envision the tone and facial expressions associated with a piece of text when it is narrated. Consequently, HyDiscGAN does not require any client-side computation during inference, reducing large collaborative costs to boost \textit{efficient communication}.
%Note that the cross-modal cGAN is pre-trained to guarantee \textit{effective multimodal alignment}, where global generators and local discriminators are distributively optimized in an alternating manner. Subsequently, components for multimodal sentiment classification are trained along with the frozen generators. 

Our key contributions can be summarized as follows:
\begin{itemize}
    \item We propose a novel hybrid DCL framework HyDiscGAN for visual-audio privacy preservation in MSA. To the best of our knowledge, this forms the first endeavor to address modality-specified privacy preservation.
    \item We customize a cross-modality cGAN to achieve effective multimodal alignment and efficient collaborative communication in HyDiscGAN.
    \item Experiments on two MSA benchmarks show HyDiscGAN achieves desirable performance competent to SOTA baselines while preserving visual-audio privacy.
\end{itemize}
\section{Related Work}
\begin{figure*}[t]
\centering
\includegraphics[height=2.6in]{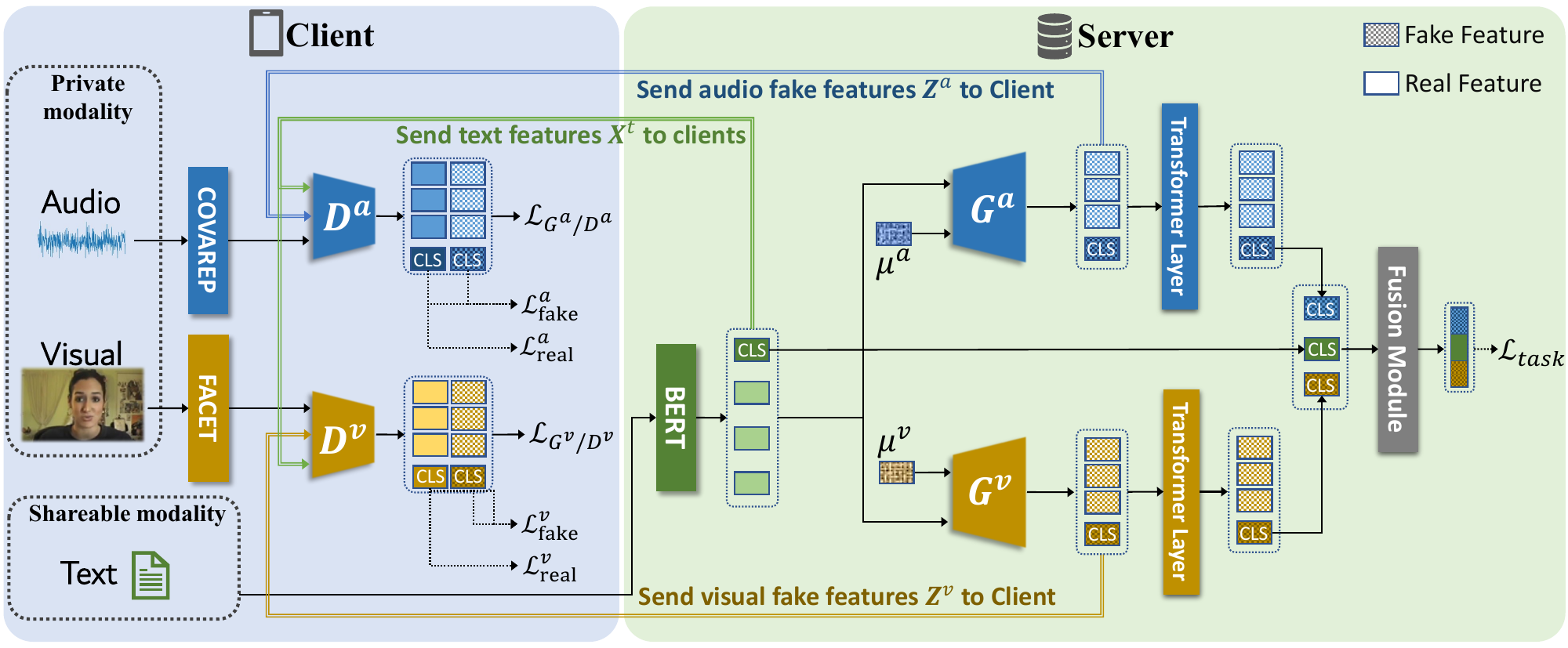}
\caption{The overall process of HyDiscGAN generates ``sufficiently realistic" fake features for private modalities (audio and visual) through hybrid DCL between the server and clients and subsequently fuses them with the real features of the shareable modality (text) for MSA tasks.}
\label{figure2}
\vspace{-2mm}
\end{figure*}
\subsection{Multimodal Sentiment Analysis}
The current approaches to MSA can be broadly categorized: representation-based and fusion-based methods~\cite{yu2023conki,LaoZSCYHM24}. Representation-based methods aim to acquire effective representations for each modality, facilitating subsequent fusion processes. One perspective argues that effective sentiment representations should encompass both modality-invariant and modality-invariant features~\cite{hazarika2020misa,yu2021learning,lin2022multimodal}. Another viewpoint suggests that, in multimodal data, the text modality predominates, seeking to enhance text representation by integrating textual and non-textual modality information~\cite{wang2019words,yang2021mtag,guo2022dynamically,SuZSLH23}. Recently, Yang et al.~\cite{yang2023confede} further improved multimodal informative representation by incorporating contrastive learning and contrastive feature decomposition alongside representation learning. Regarding fusion-based methods, early research categorized them into early-fusion and late-fusion~\cite{yu2023conki}. Early fusion emphasizes learning dependencies in multimodal sequence data~\cite{zadeh2018memory}, while late fusion initially learns independent unimodal representations, integrating them later for sentiment inference~\cite{zadeh2017tensor}. Recently, Zhao et al.~\cite{zhao2023tmmda} achieved SOTA performance in MSA by acquiring more enriched multimodal representations through data augmentation strategies on limited datasets. 
%However, all the aforementioned methods have not attempted to address the issue of personal privacy leakage in MSA.
\subsection{Distributed Collaborative Learning}
\begin{algorithm}[!t]
\caption{Training Cross-Modality cGAN.}
\begin{algorithmic}
\renewcommand{\baselinestretch}{1.0} 
\footnotesize
\STATE \textbf{Input:} Multiple training clients; training epoch $T$; audio and visual generators $G^{*}$; audio and visual global discriminators $D^{*}$\\
% \phantom{\textbf{Input:}} Multiple training clients; training epoch $T$; 
\FOR{$ i = 1 $ to $ T $}
\STATE $S$ training clients are randomly selected; \\
\STATE $\rhd$ \textit{Send global discriminators $D^{*}$ to each client} \\
\FOR{\textbf{each} client $C$ in $[S]$ \textbf{in parallel}}
\STATE $\rhd$ \textit{Send textual data $t$ to the central server}
\STATE \textbf{Server Executes}:
\STATE $X^{t}=\mathrm{BERT}\left( t \right)$; \\
\STATE $Z^{*}=G^{*}\left( \mu^{*}_{C}, X^{t};\theta_{G^{*}}\right), *\in\{a, v\}$; \\
\STATE $\rhd$ \textit{Send features $X^{t}$ and $Z^{*}$ to client $C$}
\STATE \textbf{Client Executes}:
\STATE $X^{a}=\mathrm{COVAREP}\left( a \right)$, $X^{v}=\mathrm{FACET}\left( v \right)$; \\
\STATE $\mathcal{L}_{D^{*}}, \mathcal{L}_{\texttt{real}}^{*}=D^{*}\left([X^{*}||Z^{*}], X^{t};\theta_{D^{*}}\right), *\in\{a, v\} $; \\
Update local discriminators $D^{*}$ based on $ (1-\lambda_{D}) \mathcal{L}_{D^{*}}+\lambda_{D}\mathcal{L}_{\texttt{real}}^{*}; $ \\
\STATE $\mathcal{L}_{G^{*}}, \mathcal{L}_{\texttt{fake}}^{*}=D^{*}\left(Z^{*}, X^{t};\theta_{D^{*}}\right), *\in\{a, v\} $; \\
\STATE $\rhd$ \textit{Send losses $\mathcal{L}_{G^{*}}$, $\mathcal{L}_{\texttt{fake}}^{*}$, and updated local discriminators $D^{*}$ to the central server}\\
\ENDFOR
\STATE \textbf{Server Executes}:
\STATE Update generators $G^{*}$ based on $ (1-\lambda_{G})\mathcal{L}_{G^{*}} + \lambda_{G}\mathcal{L}_{\texttt{fake}}^{*} $; \\
Update global discriminators $D^{*}$ by averaging the local discriminator parameters received from $S$ clients
\ENDFOR
\STATE \textbf{Output:} Updated audio and visual generators $G^{*}$ \\
\end{algorithmic}
\label{algorithm1}
\end{algorithm}
DCL has gained significant attention in recent years due to its data protection capabilities~\cite{kairouz2021advances}. The two most popular frameworks are federated learning~\cite{mcmahan2017communication} and split learning~\cite{gupta2018distributed}. For federated learning, the FedAvg was initially proposed by Brendan et al.~\cite{mcmahan2017communication}. In FedAvg, a complete model is trained on each local client holding data, and the locally updated models are then sent to the server for aggregation, resulting in a global model. Subsequent researchers have made further improvements, such as introducing penalty terms to address non-convex problems~\cite{li2020federated} and incorporating momentum mechanisms to enhance its convergence speed and performance~\cite{hsu2019measuring,ZhuZCA20}. The advantages of federated learning lie in the parallelization of computations across multiple clients, while its drawback is its inapplicability to scenarios where client resources are limited. On the contrary, split learning~\cite{gupta2018distributed} divides a model, such as a deep neural network, into multiple parts, and then performs computations on different devices. Thapa et al.~\cite{thapa2022splitfed} proposed the integration of federated learning and split learning, introducing federated split learning, which eliminates the inherent limitations of both frameworks.

\subsection{Generative Adversarial Networks}
Generative Adversarial Networks (GANs) were initially proposed by Goodfellow et al.~\cite{goodfellow2014generative}. Subsequently, researchers have proposed various improvements and variations~\cite{radford2015unsupervised,almahairi2018augmented,10285474}. Importantly, conditional Generative Adversarial Network (cGAN)~\cite{mirza2014conditional} introduces conditional information during the training process, enabling the generator to produce samples related to the given conditions. GANs were initially applied in computer vision to generate realistic images in a self-supervised manner and later spread rapidly to other fields~\cite{goodfellow2014generative}, for natural language processing, like text generation~\cite{zhang2017adversarial}, adversarial training~\cite{zhang2016generating}, and data augmentation~\cite{zhao2023tmmda}. 
%In this paper, we use cGAN to generate semantically aligned audio and visual features from text features, achieving positive results in MSA.
\section{Methodology}
\subsection{Problem Statement}
MSA is formulated as a binary/multi-classification or regression task for predicting sentiment labels. In contrast to all previous centralized models, our HyDiscGAN is implemented in a more realistic and secure scenario, encompassing a central server and numerous personal clients. Each client $C$ holds $N_{C}$ video clips as training or test samples. According to the Introduction, each sample comprises shareable modality data, i.e., text ($t$), as well as two private modality data, namely, audio ($a$) and visual ($v$). The raw data and extracted features of private modalities are securely maintained on their personal clients throughout the entire process.

For each sample, we obtain its real feature embedding sequences $ X^{m}=[x^{m}_{1}, x^{m}_{2}, ..., x^{m}_{L^{m}}] \in \mathbb{R}^{L^{m} \times d^{m}} $ from three modality data using BERT~\cite{kenton2019bert}, COVAREP~\cite{degottex2014covarep}, and FACET~\cite{de2011facial}, respectively. $L^{m}$ denotes the length of the sequence and $d^{m}$ is the feature dimension. $ m \in {\{t, * \}} $ and $ * \in {\{a, v\}} $ is the set of private modalities. Following BERT, we introduce $\texttt{<CLS>}$ tag features $x^{*}_{\texttt{<CLS>}}$ at the end of audio and visual feature sequences to represent the comprehensive semantics of the sequence. $x^{*}_{\texttt{<CLS>}}$ is initialized through the average pooling of all features in the sequence.

It is worth noting that our primary motivation is to generate fake feature sequences $Z^{*}=[z^{*}_{1}, z^{*}_{2}, ..., z^{*}_{L^{*}}, z^{*}_{\texttt{<CLS>}}]\in \mathbb{R}^{(L^{*}+1) \times d^{*}}$, which approximate the real features $X^{*}$ extracted from the raw audio and visual data, rather than the raw data itself. This not only reduces sentiment-irrelevant redundant computations but also applies gradient truncation to prevent adversaries from reconstructing the raw data from gradients~\cite{thapa2022splitfed}. Subsequently, $Z^{*}$ and $X^{t}$ are utilized for the training or testing of MSA models.
\subsection{Hybrid Distributed Collaborative Learning}
The training pipeline of HyDiscGAN for MSA is shown in Figure~\ref{figure2}. Specifically, its training consists of two steps:

(1) \textbf{Training Cross-Modality cGAN} involves hybrid distributed collaborative learning among numerous clients and a central server to ensure effective multimodal alignment of generators on the server. As shown in Algorithm~\ref{algorithm1}, at the beginning of each training epoch, the central server receives textual data from a group of clients. Guided by the text semantics, the generators $G^{*}$ on the server generate fake features for the audio and visual modalities, subsequently transmitted to their respective clients. Each client computes losses $\mathcal{L}_{D^{*}}$, $\mathcal{L}_{G^{*}}$, $\mathcal{L}^{*}_{\texttt{real}}$, and $\mathcal{L}^{*}_{\texttt{fake}}$ using its own real features and the received fake features, and then sends generator's losses and the local discriminator's parameters back to the server. The central server updates generators $G^{*}$ based on the received losses and updates discriminators $D^{*}$ by averaging the local parameters received from multiple clients.

(2) \textbf{Training MSA Component} begins with specific Transformer Layers to further encode the private modality fake features from the generators $G^{*}$. Subsequently, the Fusion Module is employed to combine these private modality fake features with the real features of the shareable modality, utilized for computing the MSA task loss $\mathcal{L}_{\texttt{task}}$. During this stage, the generators $G^{*}$ are fine-tuned as part of the MSA Component, while the global discriminators remain frozen.
\subsection{Cross-Modality cGAN}
% \begin{figure}[!t]
% \centering
% \includegraphics[height=2.2in, width=2.2in]{Figures/figure3.pdf}
% \caption{Structure of Cross-Modality Autoregressive Generator, where ``Add \& Norm'' represents the combination of residual connection and layer normalization, ``Feed Forward'' is the fully connected feed-forward layer.}
% \label{figure3}
% \end{figure}
cGAN~\cite{mirza2014conditional} is a variant of the Generative Adversarial Network~\cite{goodfellow2014generative} designed to enable targeted sample generation based on given conditions. It comprises a generator and a discriminator, where the generator produces samples satisfying specific conditions, while the discriminator is used to determine whether the input samples are generated by the generator or are real.

In our conception, we aim to generate the fake features of private modalities that semantically align with corresponding shareable modality features. To achieve this, we use text information, i.e., the feature sequence $X^{t}$ encoded by BERT, as the conditional input. Additionally, since private modality features are sequential data, to maintain the contextual correlation in generated fake feature sequences $Z^{*}$, we adopt an autoregressive manner to generate features at each temporal position. The generation process can be formalized as:
\begin{equation}
z^{*}_{i}=G^{*}\left ( z_{0:i-1}^{*},\ X^{t};\theta_{G^{*}}  \right )  
\end{equation}
where $\theta_{G^{*}}$ is the set of trainable parameters. Especially, $z_{0}^{*} = \mu^{*} $ and $\mu^{*} \sim \mathcal{N} (0,1)$ is a random feature vector sampled from a Gaussian distribution.
\subsubsection{Transformer Layer}\label{Component1}
In our framework, apart from the classifier and Fusion Module, all other components utilize the Transformer~\cite{vaswani2017attention} as the backbone, with distinctions solely in input and output. Transformer is an efficient neural architecture for modeling sequential data. The core computation is the Scaled Dot-Product Attention, which is defined as:
\begin{equation}\label{eq1}
\mathrm{Att}\left ( Q, K, V  \right ) = \mathrm{softmax}( \frac{QK^{T}}{\sqrt{d_{K}}} ) V
\end{equation}
where $Q$, $K$, and $V$ are obtained by linear mapping of the input feature sequence. Furthermore, the ``Multi-head'' operation~\cite{vaswani2017attention} is used to jointly focus on different parts of the input sequence on multiple subspaces, enhancing the ability to capture information.%$d_{K}$ is the dimension of $K$. 
\subsubsection{Transformer-based Autoregressive Generator}\label{Component2}
Inspired by the neural machine translation model~\cite{vaswani2017attention}, we construct a Transformer-based Autoregressive Generator. It is a simple variant of the basic Transformer Layer. Specifically, it comprises two attention structures serving different purposes: (1) Intra-modality Multi-Head Attention is employed to capture contextual relationships within the unimodal feature sequence, and corresponding $Q^{*}_{\texttt{ra}}$, $K^{*}_{\texttt{ra}}$, and $V^{*}_{\texttt{ra}}$ are all derived from the mapping of $ Z^{*}_{0:i-1} $:%in Equation~\ref{eq1} 
\begin{equation}
[Q^{*}_{\texttt{ra}},\ K^{*}_{\texttt{ra}},\ V^{*}_{\texttt{ra}} ]= [z^{*}_{0:i-1}W^{*}_{Q_{\texttt{ra}}},\ z^{*}_{0:i-1}W^{*}_{K_{\texttt{ra}}},\ z^{*}_{0:i-1}W^{*}_{V_{\texttt{ra}}}]\nonumber
\end{equation}
where $W^{*}_{Q_{\texttt{ra}}}$, $W^{*}_{K_{\texttt{ra}}}$, and $W^{*}_{V_{\texttt{ra}}}$ are parameter matrices; (2) Inter-modality Multi-Head Attention layer is used to capture cross-modality alignment information attentive to the shareable modality feature sequence $X^{t}$. Hence, $Q^{*}_{\texttt{er}}$, $K^{*}_{\texttt{er}}$, and $V^{*}_{\texttt{er}}$ are obtained as follows:
\begin{equation}
[Q^{*}_{\texttt{er}},\ K^{*}_{\texttt{er}},\ V^{*}_{\texttt{er}}] = [z_{0:i-1}^{*}W^{*}_{Q_{\texttt{er}}},\ X^{t}W^{*}_{K_{\texttt{er}}},\ X^{t}W^{*}_{V_{\texttt{er}}}]\nonumber
\end{equation}
where $W^{*}_{Q_{\texttt{er}}}$, $W^{*}_{K_{\texttt{er}}}$, and $W^{*}_{V_{\texttt{er}}}$ are parameter matrices.
\subsubsection{Transformer-based Discriminator}
For discriminators $D^{*}$, their structure is identical to generators $G^{*}$, except for the exclusion of autoregressive iteration steps. The inputs includes fake feature sequences $Z^{*}$ generated by $G^{*}$ and corresponding real feature sequences $X^{*}$. An additional binary classifier is added to the output layer to discriminate between the generated and real features. The discriminator plays a crucial role by providing feedback to the generator to enhance its ability to generate ``sufficiently realistic" fake features.
\subsection{MSA Component}\label{Component3}
We further introduce two Transformer Layers for learning deep semantic representations of non-textual modality features. Specifically, the generated fake audio and visual feature sequences $ Z^{*} $ are encoded through corresponding Transformer Layers before being fed to the Fusion Module.

% In this component, the above Attention module is employed to capture contextual relationships within the unimodal feature sequence, thus termed Intra-modality Multi-Head Attention. Specifically, the generated private modality fake feature sequences $ Z^{*} $ are further encoded through Transformer Layers corresponding to different modalities before being fed to Fusion Module. That is, in Intra-modality Multi-Head Attention, $Q^{*}$, $K^{*}$, and $V^{*}$ all come from the mapping of $ Z^{*} $:
% \begin{equation}
% Q^{*}, K^{*}, V^{*} = Z^{*}W^{Q^{*}},\ Z^{*}W^{K^{*}},\ Z^{*}W^{V^{*}}
% \end{equation}
% where $W^{Q^{*}}$, $W^{K^{*}}$, and $W^{V^{*}}$ are parameter matrices.
\subsubsection{Fusion Module}
This module is used to fuse \texttt{<CLS>} tag features of different modalities and regulate the influence of each modality feature in the final sentiment prediction via the gated attention unit~\cite{dhingra2017gated}. The operation of the gated attention unit is formulated for each modality as follows:
\begin{equation}
h_{\texttt{output}}^{m} = \mathrm{GAtt}(h_{\texttt{input}}^{m}; \theta_{\mathrm{GAtt}}^{m}) \odot h_{\texttt{input}}^{m}
\end{equation}
where the gated attention function $\mathrm{GAtt}$ is a fully connected linear layer with sigmoid activation, and its output dimension is equal to the input dimension. $\theta_{\mathrm{GAtt}}$ is the set of trainable parameters. The symbol $\odot$ denotes the Hadamard product~\cite{horn1990hadamard}. Specifically, $h_{\texttt{input}}=x_{\texttt{<CLS>}}^{t}$ for the text modality and $h_{\texttt{input}}=z_{\texttt{<CLS>}}^{*}$ for the audio and visual modalities. Finally, the tensor $h_{\texttt{final}}=$ [$ h_{\texttt{output}}^{v}:\ h_{\texttt{output}}^{t}:\ h_{\texttt{output}}^{a} $], connecting features from three modalities, is utilized for speaker's sentiment prediction.
\subsection{Learning Objectives}
Our framework contains three learning objectives:  cGAN Losses, customized Contrastive Losses, and MSA task Loss.
\subsubsection{cGAN Losses}
For a training sample that has private modality real feature sequences $X^{*}$ and generated fake feature sequences $Z^{*}$, cGAN losses are defined as:
\begin{equation}
\mathcal{L}_{G^{*}}= \frac{{\tiny 1}}{{\tiny L^{*}+1}}\sum_{i=1}^{L^{*}+1} [\mathrm{log}(1-D^{*}(G^{*}(z_{{\small 0} :i-1}^{*}, X^{t})))]
\end{equation}
\begin{equation}
\begin{split}
\mathcal{L}_{D^{*}} = \frac{{\tiny 1}}{{\tiny L^{*}+1}} & \sum_{i=1}^{L^{*}+1}[\mathrm{log}(1-D^{*}(x_{{\small 0} :i-1}^{*}, X^{t})) \\
&+\mathrm{log}D^{*}(G^{*}(z_{{\small 0} :i-1}^{*}, X^{t}))]
\end{split}
\end{equation}
where $\mathcal{L}_{G^{*}}$ and $\mathcal{L}_{D^{*}}$ represent the losses of the generator and discriminator, respectively. $L^{*}$ are lengths of feature sequences. Specifically, the \texttt{<CLS>} tag feature is also utilized for computation. In the training of Cross-modality cGAN, $\mathcal{L}_{G^{*}}$ and $\mathcal{L}_{D^{*}}$ are alternately minimized.
\subsubsection{Contrastive Losses}
We design two sample separation loss terms based on NT-Xent contrastive loss~\cite{chen2020simple}, which are used to further regularize the learning process for both the discriminator and the generator. Specifically, for a sample in training client $C$, its real and fake \texttt{<CLS>} tag features are  $x^{*}_{\texttt{<CLS>}} $ and  $z^{*}_{\texttt{<CLS>}} $, respectively. (1) the Real-Real contrastive loss $\mathcal{L}^{*}_{\texttt{real}}$ is employed to regulate the discriminator:
\begin{equation}
\mathcal{L}^{*}_{\texttt{real}} = -\mathrm{log}\frac{e^{\left(\operatorname{sim}\left(x^{*}_{\texttt{<CLS>}},\ {x^{*}_{\texttt{<CLS>}}}^{\mathbf{+}}\right) / \tau\right)}}{\underset{\{ {x^{*}_{\texttt{<CLS>}}}^{\mathbf{-}} \}\in C} \sum e^{\left(\operatorname{sim}\left(x^{*}_{\texttt{<CLS>}},\ {x^{*}_{\texttt{<CLS>}}}^{\mathbf{-}}\right) / \tau\right)}}
\end{equation}
where $\mathrm{sim}$ is Cosine similarity function and $\tau$ is the temperature parameter. $\{ {x^{*}_{\texttt{<CLS>}}}^{\mathbf{-}} \}\in C$ denotes the feature set of samples in client $C$ with a different sentiment polarity from the sample corresponding to $ x^{*}_{\texttt{<CLS>}} $. Conversely, ${x^{*}_{\texttt{<CLS>}}}^{\mathbb{+}}$ is the feature of a sample with the same sentiment polarity, randomly sampled from client $C$.

(2) the Real-Fake contrastive loss $\mathcal{L}^{*}_{\texttt{fake}}$ is introduced to regulate the generator:
\begin{equation}
\mathcal{L}^{*}_{\texttt{fake}} = -\mathrm{log}\frac{e^{\left(\operatorname{sim}\left(z^{*}_{\texttt{<CLS>}},\ x^{*}_{\texttt{<CLS>}}\right) / \tau\right)}}{\underset{\{ {z^{*}_{\texttt{<CLS>}}}^{\mathbf{other} } \}\in C} \sum e^{\left(\operatorname{sim}\left(z^{*}_{\texttt{<CLS>}},\ {z^{*}_{\texttt{<CLS>}}}^{\mathbf{other} }\right) / \tau\right)}}
\end{equation}
where $\{ {z^{*}_{\texttt{<CLS>}}}^{\mathbf{other} } \}\in C$ is the feature set of samples in client $C$, excluding the sample corresponding to $ z^{*}_{\texttt{<CLS>}} $.
\subsubsection{MSA Loss}
Let $y$ and $\hat{y}$ denote the true and predicted sentiment labels of a sample, respectively. The MSA task loss $\mathcal{L}_{\texttt{task}}$ is defined:
% \begin{equation}
% \mathcal{L}_{\texttt{task}} = \frac{1}{N_{B}}\sum_{n=1}^{N_{B}} {(y_{n}-\hat{y}_{n})}^{2}
% \end{equation}
\begin{equation}
{\mathcal{L}_{\texttt{task}}} =
\begin{cases}
{\frac{1}{N_{B}}\sum_{n=1}^{N_{B}} {y_{n} \cdot \mathrm{log} \hat{y}_{n}}}&{\text{for classification}}\\
{\frac{1}{N_{B}}\sum_{n=1}^{N_{B}} {(y_{n}-\hat{y}_{n})}^{2}}&{\text{for regression}}
\end{cases}
\end{equation}
where $N_{B}$ is the batch size. $\hat{y}$ is obtained through classification or regression predictions on $h_{\texttt{final}}$.
\section{Experiments}

\subsection{Datasets and Distributed Settings}
\begin{table*}[t]
\centering
\scalebox{0.7}{
\begin{tabular}{lrrrrrrrrrr}
\toprule
\multirow{2}{*}{\textbf{Model}} & \multicolumn{5}{c}{\textbf{MOSI}} & \multicolumn{5}{c}{\textbf{MOSEI}} \\ \cmidrule(lr){2-6} \cmidrule(lr){7-11}
                        & \multicolumn{1}{c}{Acc-2 $\uparrow$} & \multicolumn{1}{c}{F1-Score $\uparrow$} & \multicolumn{1}{c}{Acc-7 $\uparrow$} & \multicolumn{1}{c}{MAE $\downarrow$} & \multicolumn{1}{c}{Corr $\uparrow$} & \multicolumn{1}{c}{Acc-2 $\uparrow$} & \multicolumn{1}{c}{F1-score $\uparrow$} & \multicolumn{1}{c}{Acc-7 $\uparrow$} & \multicolumn{1}{c}{MAE $\downarrow$} & \multicolumn{1}{c}{Corr $\uparrow$} \\ \cmidrule(lr){1-6} \cmidrule(lr){7-11}
(G) TFN~\cite{zadeh2017tensor} & - / 80.8 & - / 80.7 & 34.9 & 0.901 & 0.698 & - / 82.5 & - / 82.1 & 51.6 & 0.593 & 0.700 \\
(G) LMF~\cite{liu2018efficient} & - / 82.4 & - / 82.4 & 33.2 & 0.917 & 0.695 & 78.5 / 81.9 & 79.0 / 81.7 & 51.6 & 0.573 & 0.714 \\
(G) MFN~\cite{zadeh2018memory} & 77.4 / - & 77.3 / - & 34.1 & 0.965 & 0.632 & 79.0 / 82.9 & 79.6 / 82.9 & 51.3 & 0.573 & 0.718 \\
(G) MulT~\cite{tsai2019multimodal} & - / 83.0 & - / 82.8 & 40.0 & 0.871 & 0.698 & - / 82.5 & - / 82.3 & 52.8 & 0.580 & 0.703 \\ \cmidrule(lr){1-6} \cmidrule(lr){7-11}
(B) MISA~\cite{hazarika2020misa} & 81.8 / 83.4 & 81.7 / 83.6 & 42.3 & 0.783 & 0.761 & \underline{83.6} / 85.5 & \textbf{83.8} / 85.3 & 52.2 & 0.555 & 0.756 \\
(B) MTAG~\cite{yang2021mtag} & - / 82.3 & - / 82.1 & - & 0.866 & 0.722 & - & - & - & - & - \\ 
(B) Self-MM~\cite{yu2021learning} & 83.4 / 85.5 & 83.4 / 85.4 & \textbf{46.7} & \underline{0.708} & \underline{0.796} & \textbf{83.8} / 85.2 & \textbf{83.8} / 84.9 & 53.9 & \underline{0.531} & \underline{0.765} \\
(B) TMMDA~\cite{zhao2023tmmda} & - / \textbf{86.9} & - / \textbf{86.9} & - & \textbf{0.703} & \textbf{0.801} & - & - & - & - & - \\
(B) ConFEDE~\cite{yang2023confede} & \textbf{84.2} / 85.5 & \textbf{84.1} / 85.5 & 42.3 & 0.742 & 0.784 & 81.7 / \underline{85.8} & \underline{82.2} / \underline{85.8} & \textbf{54.9} & \textbf{0.522} & \textbf{0.780} \\
\cmidrule(lr){1-6} \cmidrule(lr){7-11}
(B) \textbf{HyDiscGAN} (ours) & \underline{84.1} / \underline{86.7} & \underline{83.7} / \underline{86.3} & \underline{43.2} & 0.749 & 0.782 & 81.9 / \textbf{86.3} & 82.1 / \textbf{86.2} & \underline{54.4} & 0.533 & 0.761 \\
\toprule
\end{tabular}}
% \vspace{-1mm}
\caption{\label{tab:Experiments1}
Predicted results of different MSA models on MOSI and MOSEI datasets. ``$\uparrow$” indicates that larger values represent better results and ``$\downarrow$” signifies the opposite. (G) and (B) represent using Glove and BERT as text feature extractors, respectively. In Acc-2 and F1 score columns, the number on the left side of ``/” corresponds to ``negative/non-negative” and the number on the right side corresponds to ``negative/positive”. Bold values represent optimal performance and underlined values indicate suboptimal performance.
}
\end{table*}
% \begin{table}[t]
% \centering
% \scalebox{0.7}{
% \begin{tabular}{lrrrrrr}
% \toprule
% \multirow{2}{*}{Dateset} & \multicolumn{2}{c}{Train} & \multicolumn{2}{c}{Valid} & \multicolumn{2}{c}{Test} \\ \cmidrule(lr){2-3}\cmidrule(lr){4-5}\cmidrule(lr){6-7}
%                          & \multicolumn{1}{c}{\#S}          & \multicolumn{1}{c}{\#Sp}       & \multicolumn{1}{c}{\#S}          & \multicolumn{1}{c}{\#Sp}       & \multicolumn{1}{c}{\#S}         & \multicolumn{1}{c}{\#Sp}       \\  \cmidrule(lr){1-3}\cmidrule(lr){4-5}\cmidrule(lr){6-7}
% MOSI                     & 1,284        & 52           & 229          & 10           & 686         & 31           \\
% MOSEI                    & 16,326       & 150           & 1,871        & 50           & 4,659       & 100  \\ \toprule           
% \end{tabular}}
% \caption{Statistics of MOSI and MOSEI Datasets. \#S is the number of video clips, i.e., the number of samples. \#Sp is the number of distinct speakers, i.e., the number of clients.}
% \label{tab:Datasets}
% \end{table}
Two popular MSA benchmark datasets, MOSI~\cite{7742221} and MOSEI~\cite{zadeh2018multimodal}, are utilized to evaluate the performance of our HyDiscGAN. Detailed descriptions of each dataset and their corresponding distributed settings are provided in Appendix A.
% We assume that the textual data in these two datasets has been de-identified and is no longer subject to special processing. The statistics are shown in Table~\ref{tab:Datasets}.

% \textbf{MOSI} collects 2,199 video comment clips from YouTube with each clip representing a monologue by a speaker. These clips are contributed by 93 distinct speakers. The builder also considers the division according to speakers when dividing training, validation, and testing portions. The training portion comprises 52 speakers, while the validation and testing portions include 10 and 31 speakers, respectively. We naturally treat each speaker as an independent personal client with varying numbers of video clips containing sentiment scores ranging in [-3, +3]. +3 indicates the strongest positive sentiment, while -3 represents the strongest negative sentiment.

% \textbf{MOSEI} collects 22,856 video clips with sentiment scores ranging in [-3, 3]. However, the builder did not provide speaker tags. To ensure fairness, we follow previous work and further divide the training portion comprising 16,326 video clips, into 150 personal clients. The validation and testing portions consist of 50 and 100 clients, respectively. Each client has an equal number of samples to simulate the i.i.d. scenario~\cite{klaassen2001points}.
\subsection{Baselines}
\begin{table*}
\centering
\scalebox{0.65}{
\begin{tabular}{lc|ccc|c}
\toprule
\textbf{Term} & \textbf{ConFEDE} & \textbf{-FL} & \textbf{-SL} & \textbf{-SFL} & \textbf{HyDiscGAN} \\ 
\cmidrule(lr){1-6}
Privacy preservation & \ding{55} & \ding{51} & \ding{51} & \ding{51} & \ding{51} \\
Distributed computing & \ding{55} & \ding{51} & \ding{51} & \ding{51} & \ding{51} \\
Generative capacity & \ding{55} & \ding{55} & \ding{55} & \ding{55} & \ding{51} \\ 
No computations on testing clients & \ding{51} & \ding{55} & \ding{55} & \ding{55} & \ding{51} \\
\cmidrule(lr){1-6}
Client-side training & - & Parallel & Sequential & Parallel & Parallel \\
Scale of parameters (per client) & - & 109.5M & 23.9M & 23.9M &  77.8K \\
Scale of communication parameters (one epoch) & - & 109.5M$\times$2$S$ & 23.9M$\times$2$S$ & 23.9M$\times$2$S$ & 77.8K$\times$2$S$ \\
\bottomrule
\end{tabular}
}
% \vspace{-1mm}
\caption{Comparison of key attributes and training costs (on the MOSI dataset) of foundational DCL frameworks, including Federated Learning (-FL), Split Learning (-SL), Federated Split Learning (-SFL), and our HyDiscGAN. $S$ is the count of training clients in one epoch.}
% \vspace{-2mm}
\label{tab:Experiments2}
\end{table*}
% \begin{table}[t]
% \centering
% \scalebox{0.74}{
% \begin{tabular}{lccccc}
% \toprule
% \textbf{Scene} & Acc-2 $\uparrow$ & F1-Score $\uparrow$ & Acc-7 $\uparrow$ & MAE $\downarrow$ & Corr $\uparrow$ \\ \cmidrule(lr){1-6}
% -Full Disclosure & 83.2 / 84.7 &  83.2  / 84.9  & 42.4  &  0.753   & 0.775 \\ 
% -Visual Privacy & 83.6 / 86.4 & 83.3 / 86.1 & 42.9 &  \textbf{0.749}   & 0.778 \\
% -Audio Privacy & 83.1 / 85.2 & 82.8  / 85.2   &  42.2 &  0.751   & 0.777 \\ \cmidrule(lr){1-6}
% \textbf{HyDiscGAN} & \textbf{84.1} / \textbf{86.7} & \textbf{83.7} / \textbf{86.3} &  \textbf{43.2}  & \textbf{0.749} & \textbf{0.782} \\ \toprule
% \end{tabular}
% }
% \caption{Predicted results of diverse modality-specifiable privacy preservation scenes.}
% \label{tab:Experiments5}
% \end{table}
To validate the performance of the features generated by HyDiscGAN in MSA tasks, we conduct a comparison with several advanced and SOTA MSA models~\cite{zhao2023tmmda,yang2023confede}. These baseline models can be broadly categorized based on their backbone networks: (1) LSTM-based models, denoted as TFN~\cite{zadeh2017tensor}, LMF~\cite{liu2018efficient}, MFN~\cite{zadeh2018memory}, MISA~\cite{hazarika2020misa}, and Self-MM~\cite{yu2021learning}; (2) Transformer-based models, denoted as MulT~\cite{tsai2019multimodal}, TMMDA~\cite{zhao2023tmmda}, and ConFEDE~\cite{yang2023confede}. Additionally, there is a model MTAG~\cite{yang2021mtag} based on GNN. Note that only our HyDiscGAN utilizes generated private modality fake features for MSA, while all other baseline models neglect the preservation of the speaker's privacy.

To assess the MSA performance and communication costs of HyDiscGAN in distributed training, we deployed the latest MSA model ConFEDE across three widely used DCL frameworks: Federated Learning (-FL)~\cite{mcmahan2017communication}, Split Learning (-SL)~\cite{gupta2018distributed}, and Federated Split Learning (-SFL)~\cite{thapa2022splitfed}. ConFEDE and its three variants are implemented based on the codes provided by the authors. In -FL, each client trains a complete ConFEDE model using local data and then aggregates them in the central server. In -SL and -SFL, we adhere to the minimum split principle, aiming to perform as much computation as possible on the central server.
\subsection{Evaluation Criteria}
Following previous works~\cite{yang2023confede}, we evaluated the performance of our models on four metrics: Sentiment Binary Classification Accuracy (Acc-2), F1-Score, Mean Absolute Error (MAE), and Correlation Coefficient (Corr). Our results in both classification and regression experiments are reported as the averages of five different random seed runs. Detailed hyperparameter settings are included in Appendix B, and our source codes and processed datasets will be made publicly available upon acceptance.%~\footnote{\url{http://}}
\subsection{Performance Analysis}
\subsubsection{Comparison with Advanced MSA Models}
Table~\ref{tab:Experiments1} presents the comparative results between our proposed HyDiscGAN and other MSA models on MOSI and MOSEI datasets. In detail, HyDiscGAN achieves suboptimal performance across all classification metrics on the MOSI dataset. In the fundamental binary sentiment classification metrics (Acc-2 and F1-Score), the results of HyDiscGAN are, on average, only 0.325\% lower than the SOTA performance. Note that in the ``negative/positive" binary classification, HyDiscGAN ranks second, closely trailing behind the SOTA model TMMDA which incorporates data augmentation techniques. Moreover, on the MOSEI dataset, HyDiscGAN significantly outperforms ConFEDE and achieves SOTA performance in this task, exhibiting an average improvement of 0.45\% higher than the suboptimal model ConFEDE. Furthermore, in other metrics, HyDiscGAN also demonstrates competitiveness. This indicates that the private modality fake features generated by HyDiscGAN contain high-quality sentiment cues, comparable to real features.

HyDiscGAN does not exhibit the same level of performance in regression tasks, i.e., MAE and Corr, as in classification tasks. One possible reason is that the Real-Real contrastive loss can only separate samples with different sentiment polarities, lacking regularization for samples with the same sentiment polarity but differing only in intensity during the feature generation process.
\subsubsection{Comparison with Existing DCL Frameworks}
Table \ref{tab:Experiments2} presents a comparison of key attributes and training costs between our HyDiscGAN and three existing DCL frameworks that deploy the SOTA MSA model ConFEDE. Overall, the training costs on the client side are significantly reduced with HyDiscGAN (reduced by 99.93\% compared to -FL, and 99.68\% compared to -SL and -SFL). Simultaneously, HyDiscGAN possesses the capability to generate fake features for private modalities, resulting in zero costs on the client side during testing. In contrast, other DCL frameworks require the same costs during testing as in training, making HyDiscGAN more suitable for testing scenarios with completely limited resources.

The results of comparative MSA methods are outlined in Table~\ref{tab:Experiments3}. While HyDiscGAN may not surpass the centralized model ConFEDE on specific evaluation metrics, it demonstrates notable superiority over ConFEDE across all metrics when ConFEDE is deployed in existing DCL frameworks. The influence arises from the label distribution skew~\cite{zhang2022federated} in client data within existing DCL frameworks for ConFEDE. When applied to MSA tasks, HyDiscGAN follows the two-stage training approach and employs the hybrid DCL strategy exclusively during the stage of learning the private modality real feature distribution on the client. This stage involves self-supervised learning and is therefore not influenced by the distribution of sentiment labels.
% \subsection{Compatibility in Privacy and Performance}
% Table~\ref{tab:Experiments5} illustrates the performance impact of four different modality-specifiable privacy preservation scenarios on MSA tasks. Specifically,  "-Full Disclosure" denotes the scenario where privacy is not considered, and the real features of all three modalities are directly input into the MSA component. "-Visual Privacy" and "-Audio Privacy" respectively indicate scenarios where the Visual or Audio modality is individually specified as the private modality. In these cases, the privacy modality features are generated by HyDiscGAN, while the features of other modalities use the real features. It can be observed that HyDiscGAN achieves optimal performance in all scenarios, demonstrating that it attains privacy benefits without sacrificing performance.
\subsection{Ablation Study}
\begin{table}[!t]
\centering
\scalebox{0.74}{
\begin{tabular}{lccccc}
\toprule
\textbf{Model}      & Acc-2 $\uparrow$ & F1-Score $\uparrow$ & Acc-7 $\uparrow$ & MAE $\downarrow$ & Corr $\uparrow$ \\ \cmidrule(lr){1-6}
ConFEDE   &   \textbf{84.2} / 85.5    &     \textbf{84.1} / 85.5     &    42.3   & \textbf{0.742}   & \textbf{0.784} \\
-FL        &    81.4 / 81.7   &   81.3 / 81.5       &  40.7     &  0.803   & 0.721 \\
-SL       &    83.5 / 84.2   &   83.1 / 83.9       &  41.6     &  0.765   & 0.767 \\
-SFL      &   82.8 / 83.2   &    82.7 / 83.0     & 41.3     & 0.811   & 0.734  \\ \cmidrule(lr){1-6}
\textbf{HyDiscGAN}   &   84.1 / \textbf{86.7}    &  83.7 / \textbf{86.3}        &  \textbf{43.2}     & 0.749    & 0.782 \\ \toprule
\end{tabular}
}
\vspace{-2mm}
\caption{Predicted results of different DCL frameworks on the MOSI dataset, including Federated Learning (-FL), Split Learning (-SL), Federated Split Learning (-SFL), and our HyDiscGAN.}
\label{tab:Experiments3}
\end{table}
\begin{table}[t]
\centering
\scalebox{0.74}{
\begin{tabular}{lcccc}
\toprule
\textbf{Variant}      & Acc-2 $\uparrow$ & F1-Score $\uparrow$ & MAE $\downarrow$ & Corr $\uparrow$ \\ \cmidrule(lr){1-5}
Real feature (Only Audio)\textbf{$^{\dagger}$}  &   58.2    &    57.0     &   1.150  & 0.144 \\ 
Fake feature (Only Audio)  &   65.2    &   61.6       &  1.147  & 0.162 \\ 
Real feature (Only Visual)\textbf{$^{\dagger}$} &   57.4    &    57.0      &  1.160  & 0.143 \\
Fake feature (Only Visual)  &  65.3     &   65.1       &  1.139  & 0.168 \\ 
\cmidrule(lr){1-5}
cGAN loss (Only) &   85.3    &     84.9     &  0.751   & 0.778 \\ 
w/o $\mathcal{L}^{*}_{\texttt{real}}$        &   85.4    &    85.2      &  0.752   & 0.774 \\
w/o $\mathcal{L}^{*}_{\texttt{fake}}$ &  86.0     &   85.7       &  0.750   & 0.779 \\ \cmidrule(lr){1-5}
\textbf{HyDiscGAN}       & \textbf{86.7} & \textbf{86.3} & \textbf{0.749} & \textbf{0.782} \\ \toprule
\end{tabular}
}
\vspace{-2mm}
\caption{Ablation results of real/fake features for private modalities (audio and visual). \textbf{$\dagger$} indicates the results from the baseline TMMDA. ``w/o'' denotes “without”.}
\label{tab:Experiments4}
\end{table}
\subsubsection{Effects of Generated Fake Features}
Table~\ref{tab:Experiments4} upper section displays the performance of predicting sentiment tendencies using only visual or audio modality features. One observation is that in both modalities, the fake features we generated show significant performance improvements across all metrics compared to real features. This is attributed to the Cross-Modality cGAN we constructed, which generates non-textual modality features from text features. Since the text modality contains more sentiment cues, the generated features carry more sentiment information. Refer to Appendix C for a comprehensive analysis of privacy and performance compatibility experiments.

\subsubsection{Effects of Customized Contrastive Losses}
Table~\ref{tab:Experiments4} lower section demonstrates the impact of two customized contrastive losses $\mathcal{L}^{*}_{\texttt{real}}$ and $\mathcal{L}^{*}_{\texttt{fake}}$, developed by us, on the performance of MSA tasks. We observed a performance enhancement with both losses, especially with $\mathcal{L}^{*}_{\texttt{real}}$. This is because $\mathcal{L}^{*}_{\texttt{real}}$ is based on the regularization term between real features with different sentiment polarities. It promotes the aggregation of samples with the same polarity in the feature space while encouraging the separation of samples with different polarities, leading to a more distinct representation of sentiment information.
\subsection{Convergence Analysis}
\begin{figure}[t]
\centering
\subfloat[MOSI]{\includegraphics[width=1.4in]{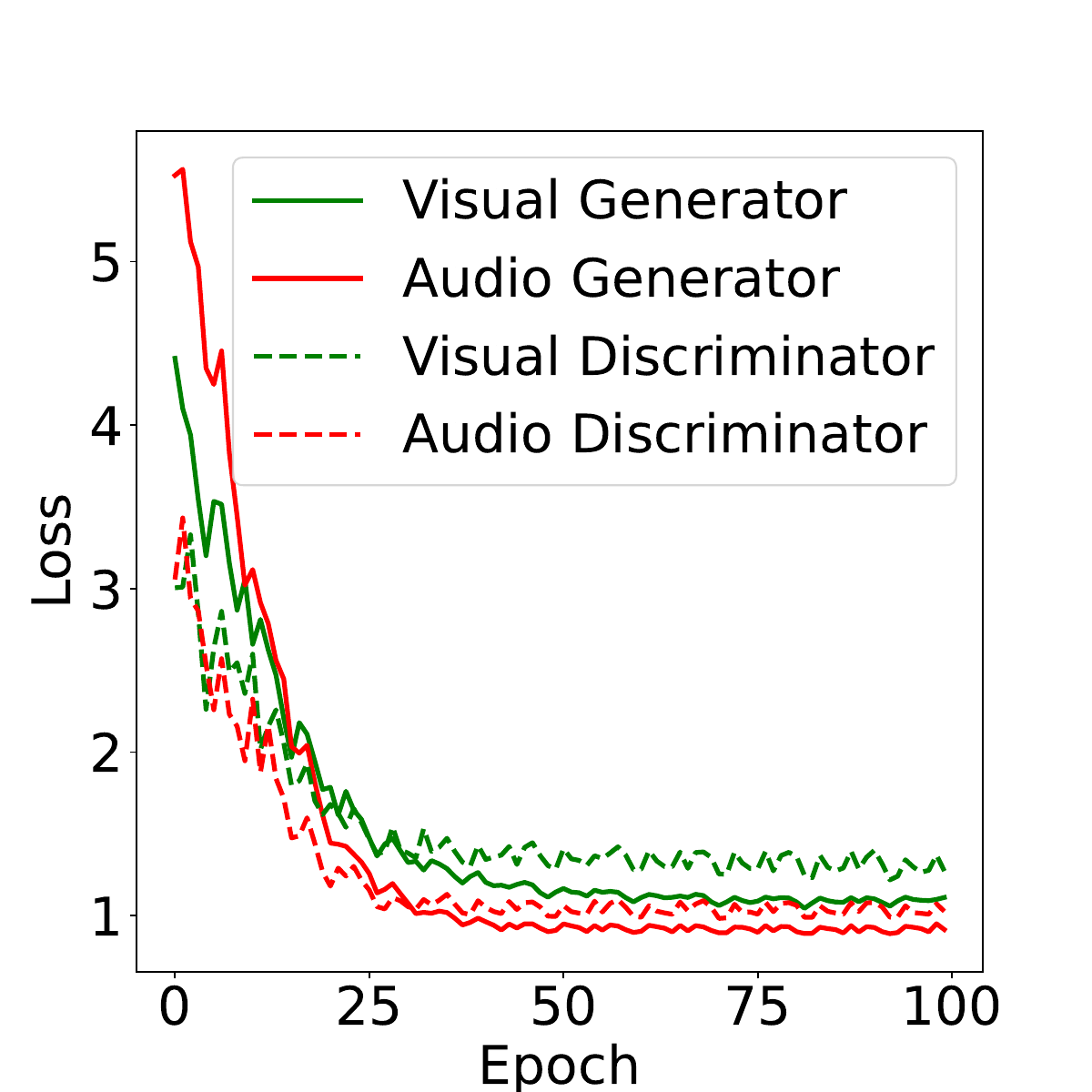}\label{line1}}
\subfloat[MOSEI]{\includegraphics[width=1.4in]{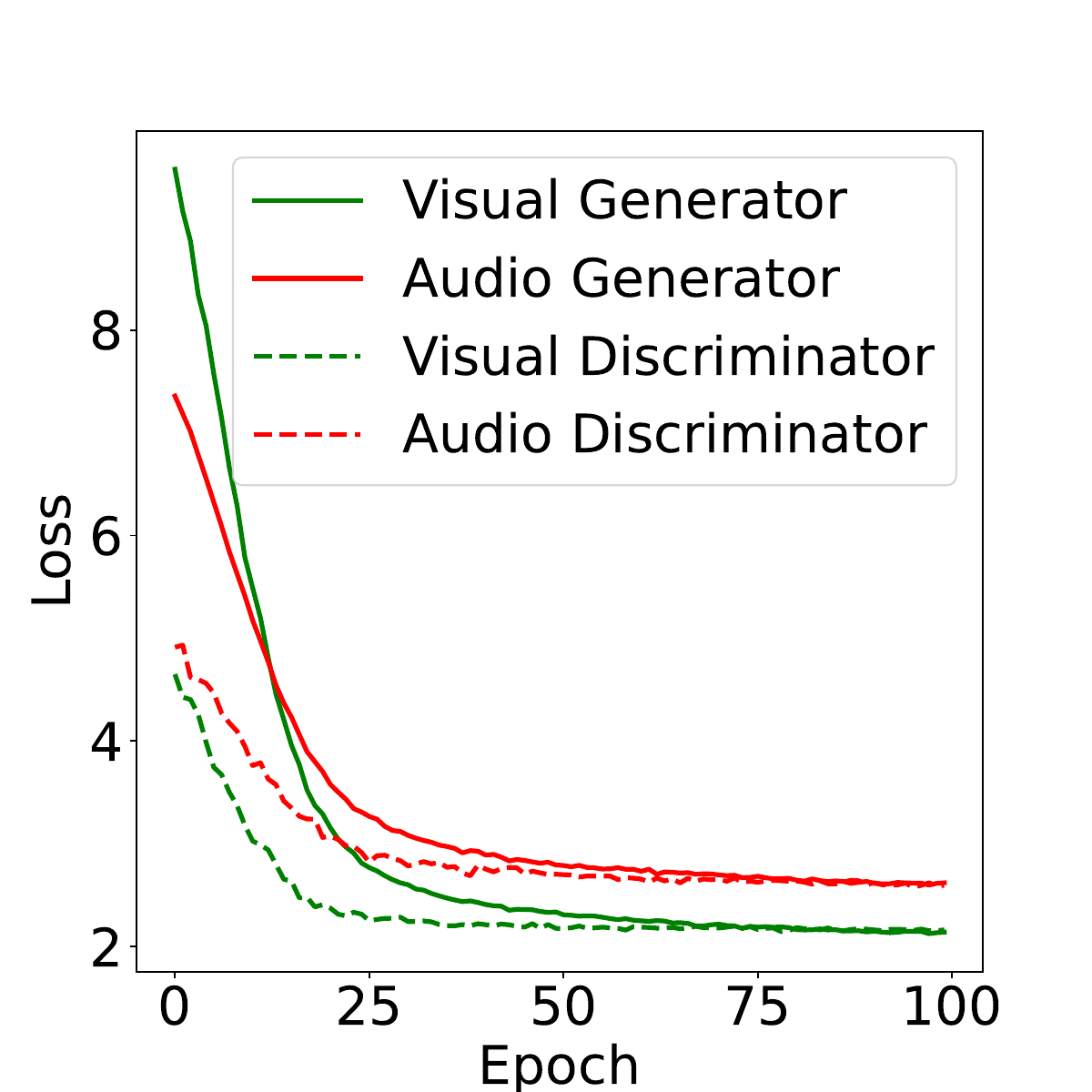}\label{line2}}
\vspace{-2mm}
\caption{Convergence visualization of training the cross-modality cGAN in HyDiscGAN on MOSI and MOSEI datasets, respectively.}
\label{figure4}
\vspace{-2mm}
\end{figure}

When training Cross-modality cGAN, there is a mutual game between the generator and discriminator, which may lead to training instability~\cite{radford2015unsupervised}. We present the convergence curves of losses during the training of Cross-Modality cGAN in HyDiscGAN on two datasets in Figure~\ref{figure4}. It can be observed that the losses of generators and discriminators eventually converge to low values. This indicates that HyDiscGAN is capable of generating ``sufficiently realistic" fake features.
% \begin{figure}[t]
%   \centering
%   \subfloat[Real visual features]{\includegraphics[width=1.2in]{Figures/MOSI_global_true_vision.pdf}\label{global1}}
%     \subfloat[Fake visual features]{\includegraphics[width=1.2in]{Figures/MOSI_global_fake_vision.pdf}\label{global2}}\qquad
%   \subfloat[Real audio features]{\includegraphics[width=1.2in] {Figures/MOSI_global_true_audios.pdf}\label{global3}}
%     \subfloat[Fake audio features]{\includegraphics[width=1.2in]{Figures/MOSI_global_fake_audios.pdf}\label{global4}}
%   \caption{t-SNE Visualization of private modality real/fake $\texttt{<CLS>}$ tag features (i.e. $x^{*}_{\texttt{<CLS>}}$ and $z^{*}_{\texttt{<CLS>}}$) for \textbf{all test samples} on the MOSI dataset, with samples from different clients labeled in different colors.}
%   \label{figure5}
% \end{figure}
\subsection{Visualization}
% \begin{figure}[t]
%   \centering
%   \subfloat[Visual features]{\includegraphics[width=1.3in]{Figures/Visualization_Visual_MOSI.pdf}\label{sample1}}
%     \subfloat[Audio features]{\includegraphics[width=1.3in]{Figures/Visualization_audio_MOSI.pdf}\label{sample2}}
%   % \subfloat[Visual\_MOSEI]{\includegraphics[width=1.6in]{Figures/Visualization_Visual_MOSEI.pdf}\label{sample3}}
%   %   \subfloat[Audio\_MOSEI]{\includegraphics[width=1.6in]{Figures/Visualization_audio_MOSEI.pdf}\label{sample4}}
%   \caption{t-SNE Visualization of private modality real/fake $\texttt{<CLS>}$ tag features for \textbf{test samples from one client} on the MOSI dataset, where positive (POS) and negative (NEG) samples are marked in red and blue, respectively.}
%   \label{figure6}
% \end{figure}
\begin{figure}[t]
  \centering
  \subfloat[Real (upper half) and fake (lower half) visual features]{\includegraphics[width=3.2in]{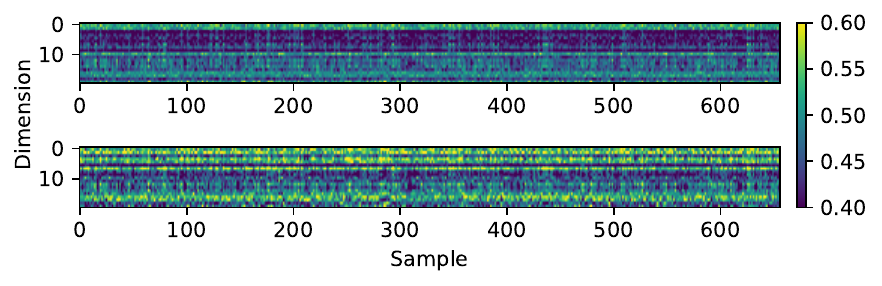}\label{hot1}}\qquad
    \subfloat[Real (upper half) and fake (lower half) audio features]{\includegraphics[width=3.15in]{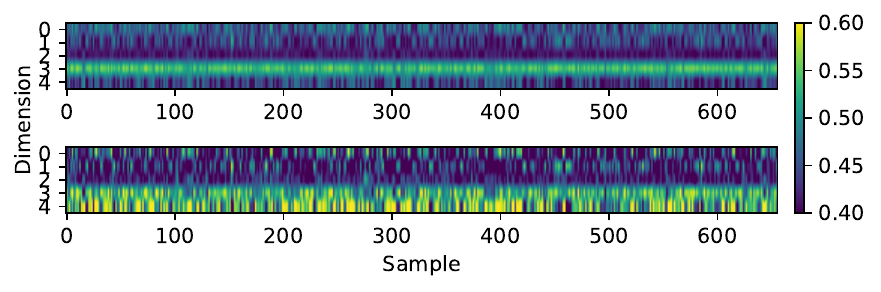}\label{hot2}}
  \caption{Visualization of the gated attention weights in the Fusion Module for visual-audio features on the test set of MOSI. Brighter regions imply higher unimodal information flow through the gates.}
  \label{figure7}
  \vspace{-2mm}
\end{figure}
To further validate the effectiveness of generated fake features for MSA tasks, we qualitatively visualize the differences in their contributions to the final sentiment prediction compared to real features. As shown in Figure~\ref{figure7}, the information in audio and visual fake features generated by HyDiscGAN is more retained in $h_{\texttt{final}}$, indicating their broader involvement in sentiment prediction and underscoring their effectiveness. More detailed visualizations are provided in Appendix D.

% Figure~\ref{figure7} depicts the difference in contribution between the generated fake features and their corresponding real features when input into the MSA Component for final sentiment prediction. 
\section{Conclusion}
In this paper, we propose a novel hybrid DCL framework HyDiscGAN for audio-visual privacy preservation in MSA. HyDiscGAN conducts training through direct communication between the server and clients, aiming to avoid constructing centralized datasets that expose personal privacy. Compared to other DCL frameworks, HyDiscGAN achieves a better balance between performance and privacy preservation. Additionally, it demonstrates significantly superior training efficiency on the client side, making it more suitable for scenarios with limited client resources. Extensive experiments verify that, while preserving privacy, HyDiscGAN competes comparably with the SOTA models in MSA tasks.

% \section*{Ethical Statement}

% There are no ethical issues.

% \section*{Acknowledgments}

% The preparation of these instructions and the \LaTeX{} and Bib\TeX{}
% files that implement them was supported by Schlumberger Palo Alto
% Research, AT\&T Bell Laboratories, and Morgan Kaufmann Publishers.
% Preparation of the Microsoft Word file was supported by IJCAI.  An
% early version of this document was created by Shirley Jowell and Peter
% F. Patel-Schneider.  It was subsequently modified by Jennifer
% Ballentine, Thomas Dean, Bernhard Nebel, Daniel Pagenstecher,
% Kurt Steinkraus, Toby Walsh, Carles Sierra, Marc Pujol-Gonzalez,
% Francisco Cruz-Mencia and Edith Elkind.

\bibliographystyle{named}
\bibliography{ijcai24}

\newpage
\appendix
\twocolumn[ % 开始跨两列的内容
\begin{center}
    \LARGE\textbf{HyDiscGAN: A Hybrid Distributed cGAN for Audio-Visual Privacy Preservation in Multimodal Sentiment Analysis}
    \vspace{10pt}
    
    \LARGE{Appendix}
    \vspace{12pt}
\end{center}
]

\section{Dataset and Distributed Settings Details}
% Two popular MSA benchmark datasets, MOSI~\cite{7742221} and MOSEI~\cite{zadeh2018multimodal}, are utilized to evaluate the performance of our HyDiscGAN. Detailed descriptions of each dataset and their corresponding distributed settings are provided in Appendix A.
In this section, we provide a detailed description of the datasets used in the experiments and their corresponding distributed settings. We assume the textual data within these two datasets has been de-identified and is no longer subject to special processing. The statistical details are shown in Table~\ref{tab:Datasets}.
\begin{table}[h]
\centering
\scalebox{0.75}{
\begin{tabular}{lrrrrrr}
\toprule
\multirow{2}{*}{Dateset} & \multicolumn{2}{c}{Train} & \multicolumn{2}{c}{Valid} & \multicolumn{2}{c}{Test} \\ \cmidrule(lr){2-3}\cmidrule(lr){4-5}\cmidrule(lr){6-7}
                         & \multicolumn{1}{c}{\#S}          & \multicolumn{1}{c}{\#Sp}       & \multicolumn{1}{c}{\#S}          & \multicolumn{1}{c}{\#Sp}       & \multicolumn{1}{c}{\#S}         & \multicolumn{1}{c}{\#Sp}       \\  \cmidrule(lr){1-3}\cmidrule(lr){4-5}\cmidrule(lr){6-7}
MOSI                     & 1,284        & 52           & 229          & 10           & 686         & 31           \\
MOSEI                    & 16,326       & 150           & 1,871        & 50           & 4,659       & 100  \\ \toprule           
\end{tabular}}
\caption{Statistics of the MOSI and MOSEI Datasets. \#S represents the number of video clips, i.e., the number of samples. \#Sp indicates the number of distinct speakers, i.e., the number of clients.}
\label{tab:Datasets}
\end{table}

\textbf{MOSI} collects 2,199 video clips from YouTube with each clip representing a monologue by a speaker. These clips are contributed by 93 distinct speakers. The builder also considers the division according to speakers when dividing training, validation, and testing portions. The training portion comprises 52 speakers, while the validation and testing portions include 10 and 31 speakers, respectively. We naturally treat each speaker as an independent personal client with varying numbers of video clips containing sentiment scores ranging in [-3, +3]. +3 indicates the strongest positive sentiment, while -3 represents the strongest negative sentiment.

\textbf{MOSEI} collects 22,856 video clips with sentiment scores ranging in [-3, 3]. However, the builder did not provide speaker tags. To ensure fairness, we follow previous work and further divide the training portion comprising 16,326 video clips, into 150 personal clients. The validation and testing portions consist of 50 and 100 clients, respectively. Each client has an equal number of samples to simulate the i.i.d. scenario~\cite{klaassen2001points}.
\section{Hyperparameters}
All our models are based on Python 3.8.18 and PyTorch 2.0.0, while training and testing were conducted on a single Tesla V100 PCIe 32GB GPU. We utilize the Adam algorithm to optimize the objective losses in both stages of training (Training Cross-Modality cGAN and Training MSA Component). The key hyperparameters used in the experiments are detailed shown in Table~\ref{Hyperp}.
\begin{table}[t]
\centering
\scalebox{0.75}{
\begin{tabular}{lcc}
\toprule
\textbf{Hyperparameters}      & \textbf{MOSI} & \textbf{MOSEI} \\ \cmidrule(lr){1-1}\cmidrule(lr){2-2}\cmidrule(lr){3-3}
Num of Transformer Layers\ /\ heads  &      &       \\ 
-Visual Generator $G^{v}$  &   2\ /\ 2   &   4\ /\ 2    \\
-Audio Generator $G^{a}$  &   1\ /\ 1   &   5\ /\ 3    \\
-Visual Discriminator $D^{v}$ &   2\ /\ 2   &    4\ /\ 2    \\
-Audio Discriminator $D^{a}$ &   1\ /\ 1   &  5\ /\ 3     \\
-Visual Transformer Layer  &   2\ /\ 2   &   2\ /\ 2    \\
-Audio Transformer Layer  &   1\ /\ 1   &   2\ /\ 3    \\ \cmidrule(lr){1-1}\cmidrule(lr){2-2}\cmidrule(lr){3-3}
Feature Dimension  &      &       \\ 
-Text $x^{t}$  &   768   &    768   \\ 
-Visual $x^{v}$  &   20   &   35    \\ 
-Audio $x^{a}$  &   5   &    74   \\   \cmidrule(lr){1-1}\cmidrule(lr){2-2}\cmidrule(lr){3-3}
Learning Rate  &      &       \\
-Generator $G^{*}$  &  2e-4    &  2e-4     \\ 
-Discriminator $D^{*}$  &   1e-4   &   1e-4    \\ 
-MSA task  &   1e-4   &   1e-4    \\ 
-$\lambda_{D}$  &   0.1   &   0.1    \\ 
-$\lambda_{G}$  &   0.1   &    0.1   \\   \cmidrule(lr){1-1}\cmidrule(lr){2-2}\cmidrule(lr){3-3}
epoch $T $ in training Cross-Modality cGAN  &  100    &  100     \\ 
Randomly select the number of training clients $S$  &   10   &  5     \\ 
Batch size $N_{B}$ in training MSA Component &   32   &   32    \\ 
\toprule
\end{tabular}
}
\caption{Hyperparameters of HyDiscGAN applied to different datasets.}
\label{Hyperp}
\end{table}
\section{Compatibility in Privacy and Performance}
\begin{table}[t]
\centering
\scalebox{0.71}{
\begin{tabular}{lccccc}
\toprule
\textbf{Scene} & Acc-2 $\uparrow$ & F1-Score $\uparrow$ & Acc-7 $\uparrow$ & MAE $\downarrow$ & Corr $\uparrow$ \\ \cmidrule(lr){1-6}
-All shareable & 83.2 / 84.7 &  83.2  / 84.9  & 42.4  &  0.753   & 0.775 \\
-Audio privacy & 83.1 / 85.2 & 82.8  / 85.2   &  42.2 &  0.751   & 0.777 \\
-Visual privacy & 83.6 / 86.4 & 83.3 / 86.1 & 42.9 &  \textbf{0.749}   & 0.778 \\ 
-Audio-Visual privacy & \textbf{84.1} / \textbf{86.7} & \textbf{83.7} / \textbf{86.3} &  \textbf{43.2}  & \textbf{0.749} & \textbf{0.782} \\ \toprule
\end{tabular}
}
\caption{Prediction results of HyDiscGAN under diverse modality-specified privacy preservation scenarios on the MOSI dataset.}
\label{tab:Experiments5}
\end{table}
\begin{figure*}[t]
  \centering
  \subfloat[Audio features]{\includegraphics[height=1.3in]{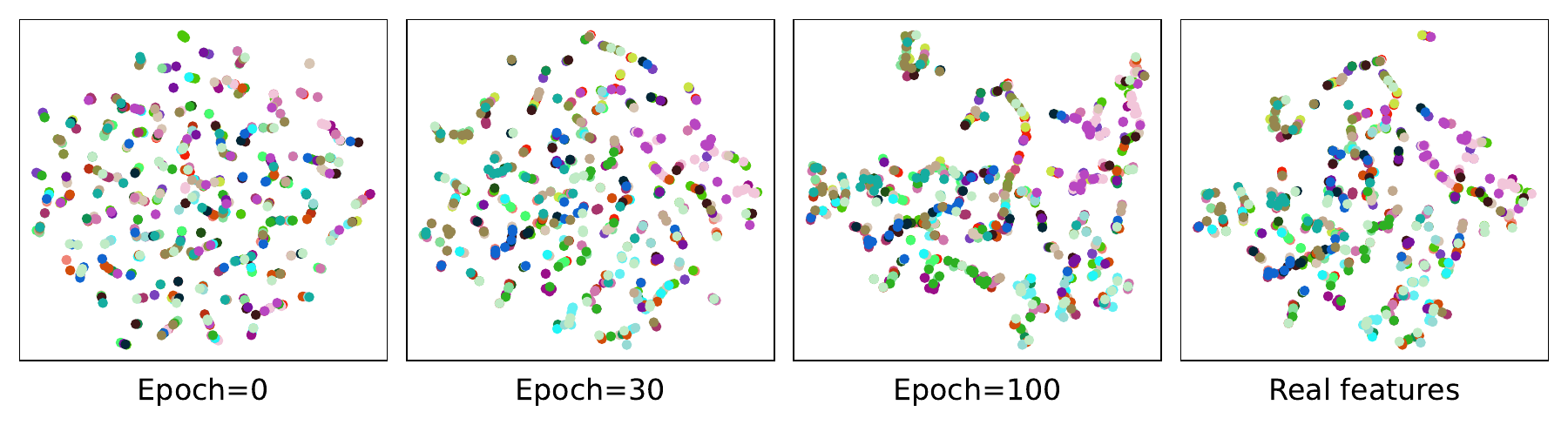}\label{global2}}\qquad
      \subfloat[Visual features]{\includegraphics[height=1.3in]{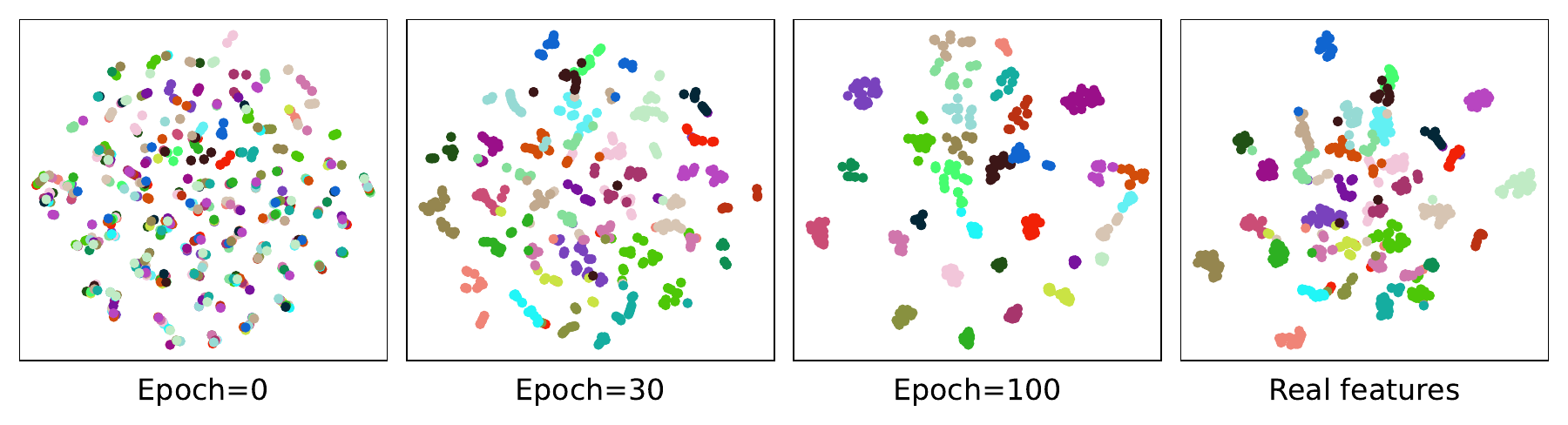}\label{global1}}
  \caption{t-SNE Visualization of private modality real/fake $\texttt{<CLS>}$ tag features (i.e. $x^{*}_{\texttt{<CLS>}}$ and $z^{*}_{\texttt{<CLS>}}$) for \textbf{all test samples} on the MOSI dataset, with samples from different clients labeled in different colors. “Epoch=0, 30, and 100” represent the fake features at different epochs during the training of Cross-Modality cGAN.}
  \label{figure5}
\end{figure*}
HyDiscGAN aims to generate fake features of privacy modalities from shareable de-identified textual data, replacing real features for downstream tasks. In practical applications, more flexible modality-specified privacy preservation scenarios may arise. Table~\ref{tab:Experiments5} presents a performance comparison in four different modality-specified privacy preservation scenarios for MSA tasks. Specifically, ``-All shareable" denotes a scenario where privacy is not considered, and real features of all three modalities are directly input into the MSA Component. ``-Audio privacy" and ``-Visual privacy" indicate scenarios where the audio or visual modality is individually specified as the private modality, respectively. ``-Audio-Visual privacy" represents the scenario where both modalities are private. In these scenarios, HyDiscGAN generates privacy modality features, while the features of other modalities use real features. Initially, it is observed that ``-Audio-Visual privacy" achieves optimal performance across all scenarios. Furthermore, the performance in ``-Audio privacy" and ``-Visual privacy" scenarios is generally superior to that in the ``-All shareable" scenario. This is attributed to the joint optimization of cGAN losses and customized contrastive losses by HyDiscGAN, enabling HyDiscGAN to generate ``sufficiently realistic" fake features and obtain clearer sentiment tendencies in generated fake features. By customizing constraints in the learning process, we can acquire fake features better suited for the target task than real features, achieving dual benefits in privacy and performance. In addition, HyDiscGAN achieves competent performance under different scenarios, indicating the applicability and flexibility of HyDiscGAN in various modality-specified privacy preservation scenarios of MSA tasks.
%provides a paradigm for other downstream tasks
% become clearer, under the constraint of the contrastive loss
% Table~\ref{tab:Experiments5} illustrates the performance impact of four different modality specifiable privacy preservation scenarios on MSA tasks. Specifically,  "-All shareable" denotes the scenario where privacy is not considered, and the real features of all three modalities are directly input into the MSA Component. "-Visual privacy" and "-Audio privacy" indicate scenarios where the Visual or Audio modality is individually specified as the private modality, respectively. In these cases, the privacy modality features are generated by HyDiscGAN, while the features of other modalities use the real features. It can be observed that HyDiscGAN achieves optimal performance in all scenarios, demonstrating that it attains privacy benefits without sacrificing performance.
\section{More visualization}
\subsection{Global Distribution of Real/Fake Features}
Figure~\ref{figure5} showcases the distribution of real/fake features for private modalities in \textbf{all test samples} from the MOSI dataset. An important observation is that, for the visual modality, real features naturally exhibit a clear clustered distribution consistent with the division of clients (i.e., speakers). With the continuous iterations of training, the distribution of fake features generated by HyDiscGAN also shows this trend. Moreover, when real features do not exhibit a clear clustered distribution, such as the audio modality, HyDiscGAN can also effectively capture the distribution of real features and preserve this characteristic. Consequently, HyDiscGAN demonstrates an excellent ability to learn the global distribution of features across different modalities.
% An important observation is that real visual features exhibit clear clusters, aligning with the division of clients (i.e., speakers), while real audio features do not show such distinct separation. This difference is effectively captured by HyDiscGAN, where the generated fake features exhibit distributional characteristics similar to the real features, demonstrating the outstanding learning capabilities of HyDiscGAN.
\subsection{Local Distribution of Real/Fake Features}
\begin{figure}[H]
  \centering
    \subfloat[Audio features (MOSI)]{\includegraphics[width=1.5in]{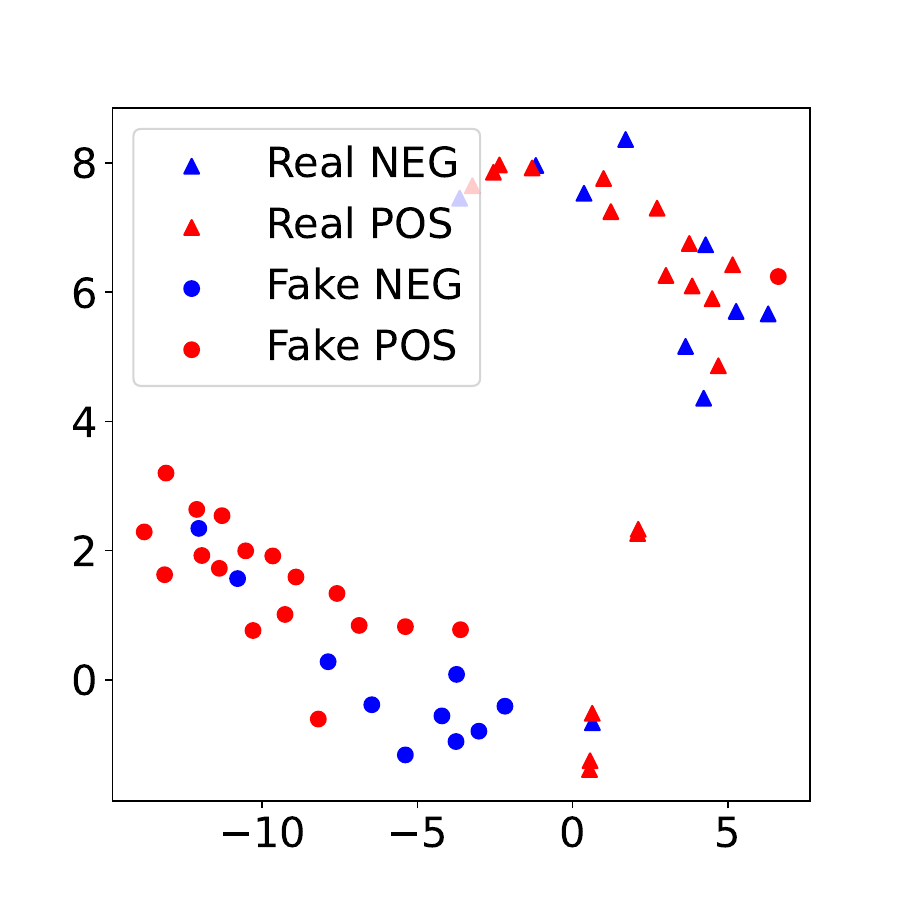}\label{sample2}}
    \subfloat[Visual features (MOSI)]{\includegraphics[width=1.5in]{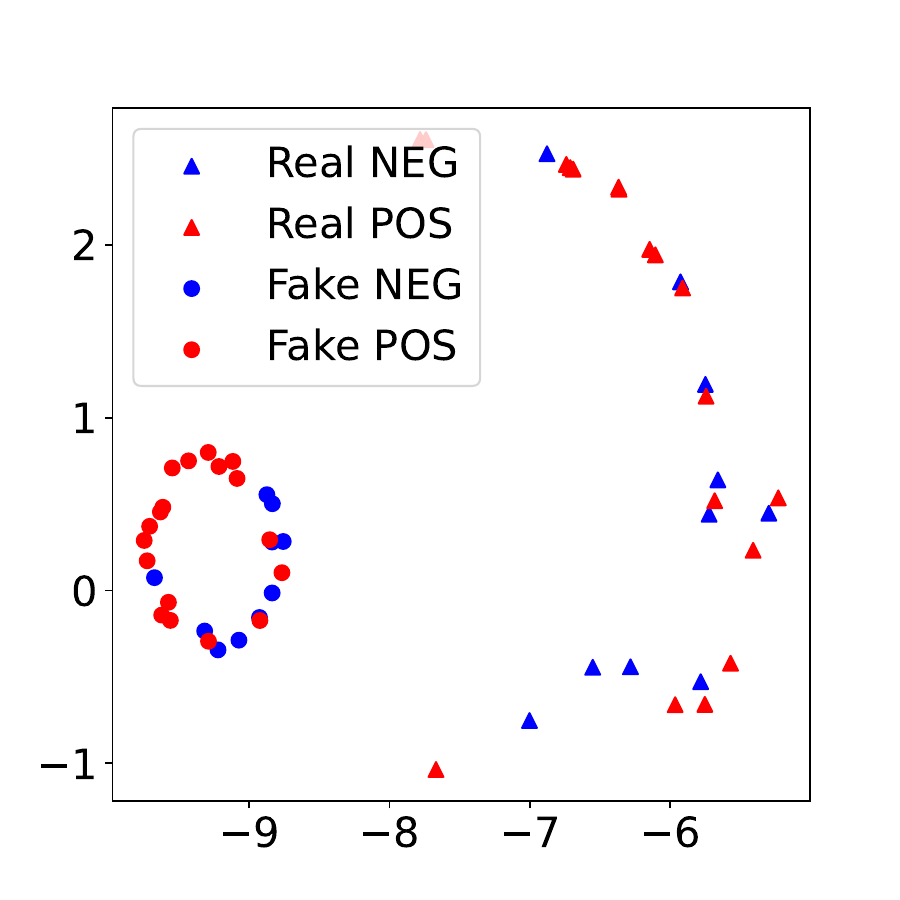}\label{sample1}}\qquad
    \subfloat[Audio features (MOSEI)]{\includegraphics[width=1.5in]{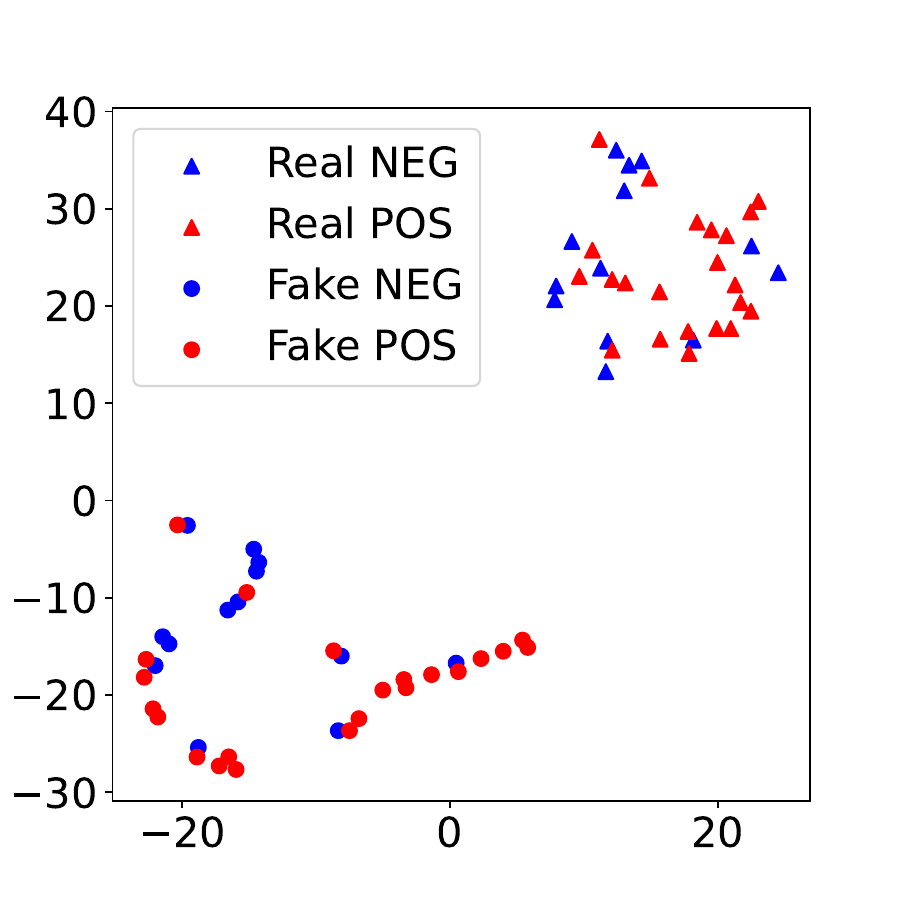}\label{sample4}}
    \subfloat[Visual features (MOSEI)]{\includegraphics[width=1.5in]{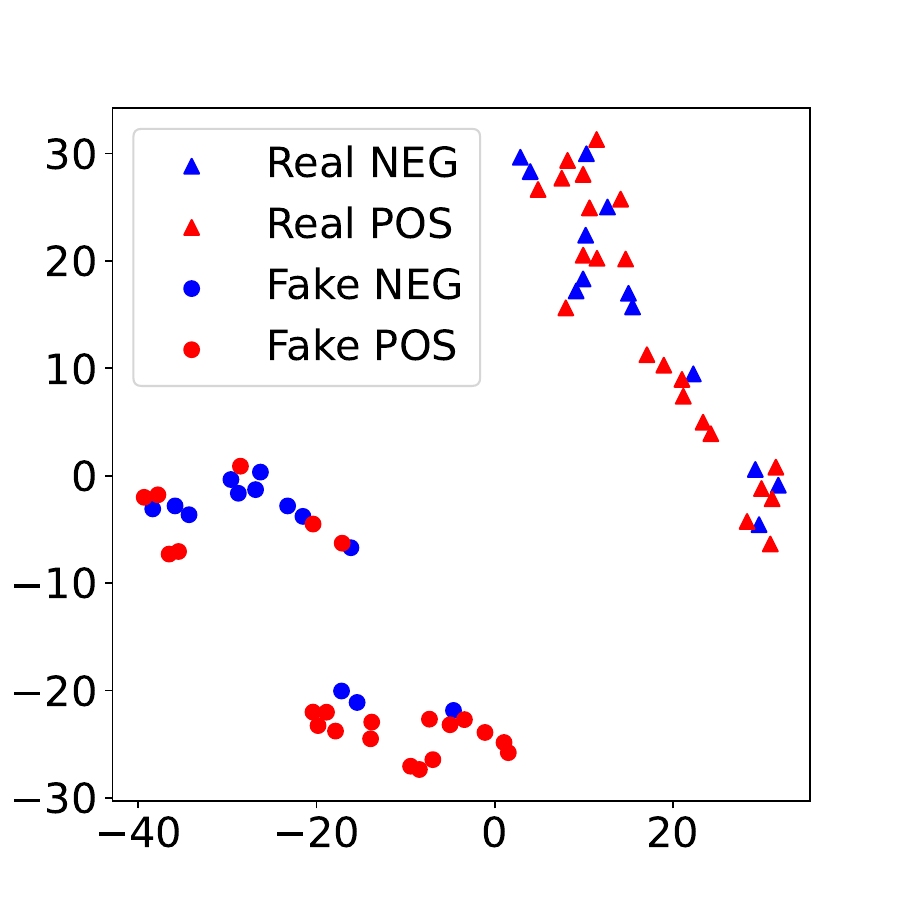}\label{sample3}}
  \caption{t-SNE Visualization of private modality real/fake $\texttt{<CLS>}$ tag features for \textbf{test samples from one client} on MOSI or MOSEI dataset, where positive (POS) and negative (NEG) samples are marked in red and blue, respectively.}
  \label{figure6}
\end{figure}

Figure~\ref{figure6} illustrates the distribution of real/fake features for private modalities in \textbf{test samples from one client} on the MOSI or MOSEI dataset. An observation reveals that the audio and visual fake features of different samples exhibit clustering tendencies based on the samples' sentiment polarity. This trend arises from the customized contrastive losses applied to various samples during the training of Cross-Modality cGAN. This observation also explains the enhanced performance of HyDiscGAN in sentiment classification.
\section{Impact of hyperparameters \texorpdfstring{$\lambda_{D/G}$}{lambda}}
\begin{figure}[t]
  \centering
      \subfloat[Classification task]{\includegraphics[height=1.5in]{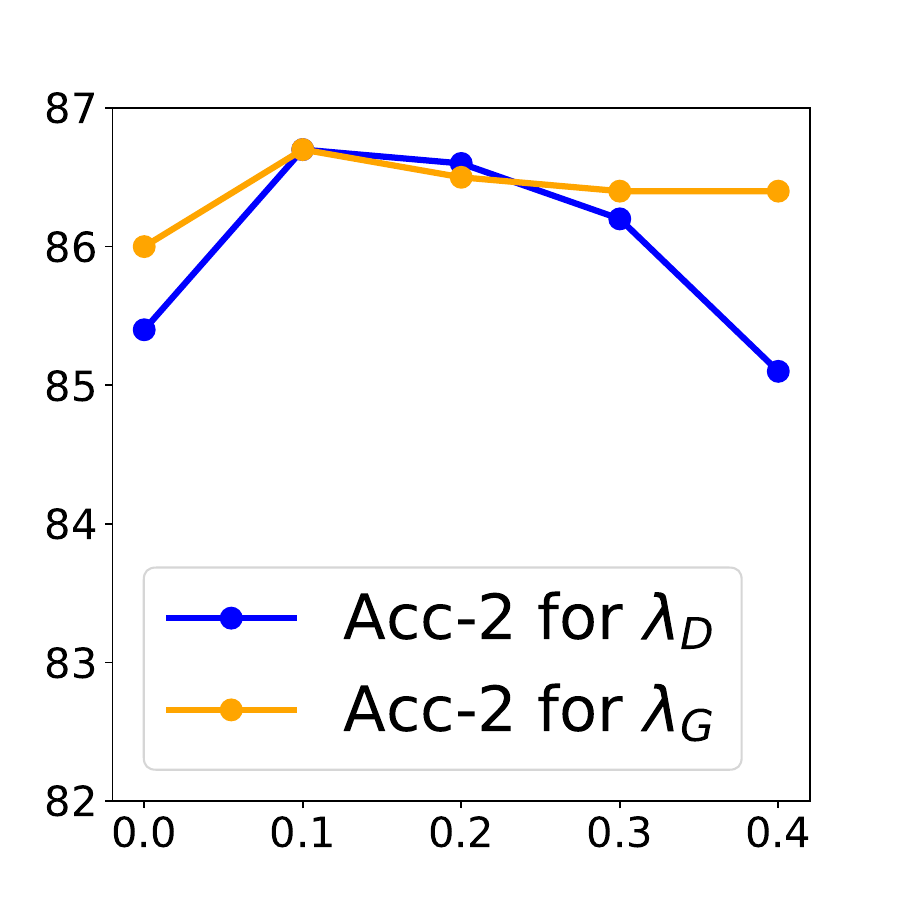}\label{para1}}\qquad
  \subfloat[Regression task]{\includegraphics[height=1.5in]{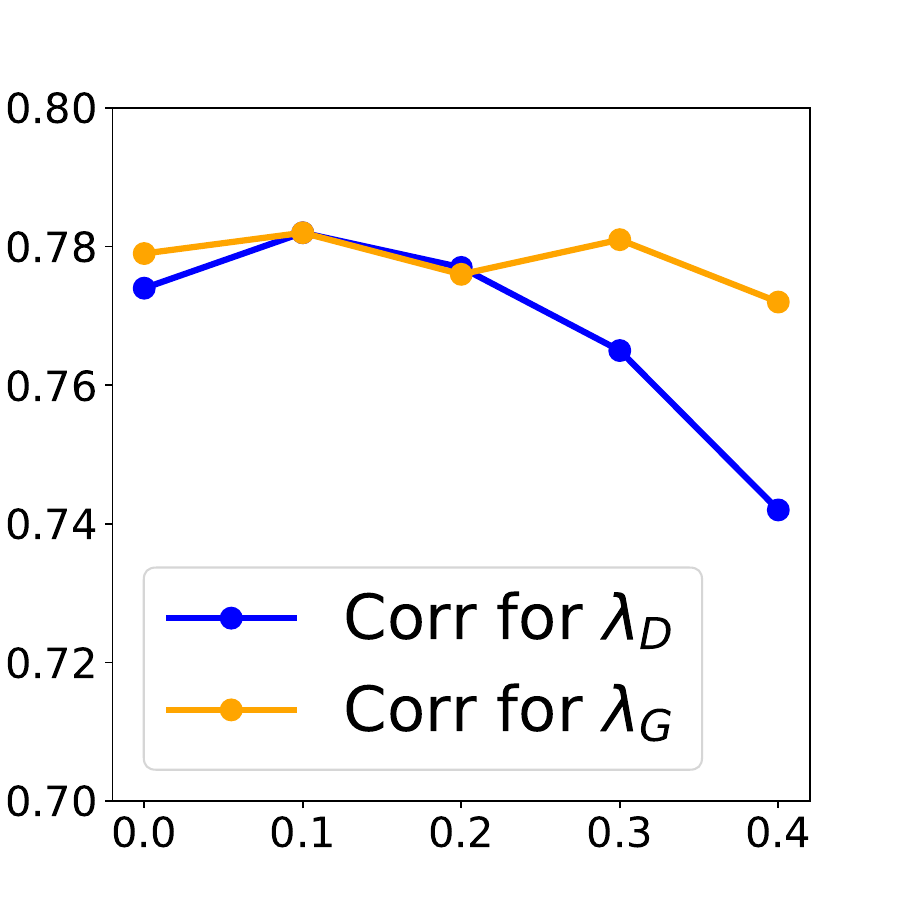}\label{para2}}
  \caption{Prediction results of MSA tasks when the hyperparameters $\lambda_{D}$ and $\lambda_{G}$ vary within the range of 0.1 to 0.4. The classification task in (a) refers to the "negative/positive" binary classification.}
  \label{figure8}
\end{figure}

The hyperparameters $\lambda_{D}$ and $\lambda_{G}$ are used to adjust the ratio of cGAN losses and customized contrastive losses during the training stage of Cross-modality cGAN. Figure~\ref{figure8} depicts their influence on the performance of the final MSA tasks. The results show the same trend in both classification and regression tasks, i.e. an initial performance increase followed by a subsequent decrease in performance. Specifically, when $\lambda_{D}$ and $\lambda_{G}$ are set to 0.1, the performance of both MSA tasks reaches optimal. Moreover, it is evident that with an increase of $\lambda_{D}$, the performance of the regression task drops sharply. This aligns with the explanation provided in the Performance Analysis, that is, Real-Real contrastive loss $\lambda_{D}$ has a limited predictive effect on sentiment intensity. Therefore, by setting both $\lambda_{D}$ and $\lambda_{G}$ to 0.1, HyDiscGAN achieves a balance between learning the distribution of real features and acquiring a clearer representation of sentiment features.
\section{Baseline Details}
\subsection{MSA Models}
\begin{itemize}
    \item TFN~\cite{zadeh2017tensor} proposes a Tensor Fusion Network for learning inter-modality and intra-modality dynamics.
    \item LMF~\cite{liu2018efficient}, known as Low-rank Multimodal Fusion network, stands as an advanced variant of the TFN. It effectively reduces the computational complexity of multimodal tensors.
    \item MFN~\cite{zadeh2018memory} is a Memory Fusion Network, that constitutes a multi-view sequential learning architecture employing attention mechanisms to achieve cross-modality interaction learning.
    \item MulT~\cite{tsai2019multimodal} utilizes a cross-modality transformer to achieve the translation from the source modality to the target modality, thereby comprehending the deep semantics of different modalities.
    \item MISA~\cite{hazarika2020misa} maps features from different modalities to distinct feature spaces, facilitating the learning of modality-invariant and modality-specific representations, thereby enhancing the capture of commonalities and differences across various modalities.
    \item MTAG~\cite{yang2021mtag} stands as the sole method employing graph neural networks to model multimodal data. It transforms multimodal data into a graph structure, capturing rich semantics across modalities and time through information aggregation on the graph.
    \item Self-MM~\cite{yu2021learning} devises a unimodal label generation network based on self-supervised learning strategies to acquire unimodal label information. Subsequently, it jointly trains unimodal and multimodal sentiment analysis, employing an adaptive weight adjustment strategy to balance progress across different tasks.
    \item TMMDA~\cite{zhao2023tmmda} proposes a token mixup technique for multimodal data augmentation, aiming to acquire efficient multimodal representations on limited labeled datasets.
    \item ConFEDE~\cite{yang2023confede} proposes a unified learning framework for contrastive representation learning and contrastive feature decomposition, aiming to acquire well-rounded multimodal information representations, including modality-invariant and modality-specific components.
\end{itemize}
\subsection{DCL Frameworks}
\begin{itemize}
    \item FL~\cite{mcmahan2017communication} introduces Federated Learning (FL) and the Federated Averaging (FedAvg) Algorithm. It trains complete models on individual clients with their data and aggregates the updates on the server to learn a global model.
    \item SL~\cite{gupta2018distributed} (Split Learning) divides the AI model and trains partial models on both the server and clients with data. In contrast to FL, it doesn't need complete model training on clients, making it suitable for resource-limited scenarios. However, client-side training cannot be parallelized.
    \item SFL~\cite{thapa2022splitfed} (SplitFed Learning) combines FL and SL methods, eliminating their respective limitations.
\end{itemize}
\end{document}